\documentstyle[preprint,epsfig,eqsecnum,aps]{revtex}

\begin{document}
\draft
\title{Unified description of three positive and three negative parity interacting bands}

\author{A. A. Raduta$^{a,b,d)}$ \footnote{Senior Fellow of Humboldt Foundation, supported within the
Programme {\it Stability Pact in South-East Europe}}, D. Ionescu $^{c,d)}\footnote{Fellow of Humboldt Foundation, supported within the Programme
{\it Stability Pact in South-East Europe}} $ and Amand Faessler$^{d)}$}

\address{
$^{a)}$Theoretical Physics and Mathematics, Bucharest University,
POB MG6, Romania}

\address{$^{b)}$Institute of Physics and Nuclear Engineering, Bucharest POB MG6, Romania}

\address{$^{c)}$Dep. of Biophysics, Medical University  "C. Davila", 
 76241, Bucharest, Romania }

\address{$^{d)}$Institute for Theoretical Physics, Tuebingen University, Auf der Morgenstelle 14, Germany} 

\date{\today}

\maketitle

\begin{abstract}
The coherent state model (CSM) is extended so that three negative parity bands are treated on equal footing with three positive parity bands. The six rotational bands are generated by projecting out  angular momenta and parities from three intrisic orthogonal states which exhibit both quadrupole and octupole deformations. The projected states are, by construction, mutually orthogonal. In the space of projected states, a sub-space of a quadrupole and octupole multi-boson states, an effective quadrupole and octupole boson Hamiltonian is solved.
The eigenstates of the model Hamiltonian are linked by multipole transition operators for which  lowest order  boson expressions are considered.
The calculations involve 6 structure coefficients and two deformation parameters. All of them are fixed by a least square fit of the known experimental energies. Applications are made to $^{158}$Gd, $^{172}$Yb, $^{218}$Ra, $^{226}$Ra, $^{232}$Th, $^{238}$U, $^{238}$Pu. Very good agreement is obtained for both excitation energies and transition probabilities. New signatures for octupole deformation, manifested in excited bands, are pointed out.
\end{abstract}


\section{Introduction}
\label{sec:level1}
The field of negative parity states is of about the same age as the field concerned with the positive parity states. The pioneering papers in this domain appeared already in the beginning of the fifties\cite{Asar,Step} and were based on high resolution alpha spectroscopy measurements. The states  were identified
 as $1^-,3^-,5^-$ through angular correlations and conversion coefficients analysis, as well by measuring the E1 branching ratio for the first state to $0^+$ and $ 2^+$. Also, it was concluded that these states belong to a $K^{\pi}=0^-$ band.
Since that early time, a lot of work has been  devoted to this issue from both experimental and theoretical sides
\cite{Stru,Bo,Dav,Dom,Lip,Rad1,Neer1,Neer2,Rad2,Rad3,Rad4,Nom,Cea,Kam,Rad5,Cha1,Cha2,Kam1,Mol,Peck,Sol1,Lea,Roh1,Alh,Iach,Dal,Rad6,Wa,Gai,Da,Naza,Ah,Cel,Dal1,Roho,Sobi,Barf,Cot,Ham,Rad7,Dass,Ahmad,ButNaz,Sug,Blo,Lee,Gre,Bur,Bal,Gai1,Coc1,Ale,Shu,Sim,Gro,Wol,Ell,Gre1,Gre2,Smi,Led,Coc2,Han,Kho,But,Bem,McG}. In the early stage of investigation, the negative parity states were considered as pure octupole vibrational states. Thus in a phenomenological language they are determined by small oscillations of the nuclear surface parametrized by quadrupule and octupole shape variables around a
spherical equilibrium shape. In the framework of a microscopic descriptions they are octupole particle hole RPA (random phase approximation) states.  

The interest in the field of negative parity states increased considerably when first suggestions for octupole deformed nuclei appeared. Indeed, in Ref.\cite{Cha1,Cha2} Chassman predicted parity doublets for several odd mass isotopes of Ac, Th and Pa. The doublet members have the same angular momentum, about the same energy but different parities. The lowest doublet constitutes a degenerate ground state which is due to a reflection symmetry for the equilibrium shape. If the doublet is not mathematically degenerate but exhibits a small energy split, this is a sign of a reflection symmetry breaking. 
This assumption allowed for a consistent description of all available data  for the low lying spectra of these isotopes.
Another theoretical work,  which suggested that some even-even isotopes of Ra isotopes might have an octupole deformed ground band, belongs to Moller and Nix
\cite{Mol}. Their calculations showed  that the binding
 energy of these nuclei gains about 2 MeV when in the mean field an octupole deformation is assumed.

The difficulty in the experimental study of pear shaped nuclei, is a missing
 observable which might be interpreted as a measure for the octupole deformation. Therefore some indirect information about this variable should be found.
Along the time several signatures were assigned to the octupole deformation:
a) The low position of the state $1^-$ heading the band $0^-$ is an indication that, as function of the octupole deformation, the potential energy has a flat minimum.
b) The parity alternating structure in ground and the low $0^-$ bands suggests that the two bands may be viewed as being projected from a sole deformed intrinsic state, exhibiting both quadrupole and octupole deformations. If that is the case, the moments of inertia in the two bands are the same which results in obtaining a vanishing displacement energy function (see section 5).
c)  A nuclear surface with quadrupole and octupole deformations might have the center of charge in a different position than the center of mass, which results in having a dipole moment which may excite the state $1^-$ from the ground state, with a large probability.

The main achievements in the field of negative parity bands have been reviewed in two recent papers \cite{Ahmad,But}. In order to save space we shall not dwell on the history of this field and advise the reader to consult the two 
papers quoted above.

In refs.\cite{Rad8,Rad9,Rad10} two authors  of the present paper (A.A.R and A.F) proposed a phenomenological model to describe simultaneously  the ground band and $K^{\pi}=0^-$ band. The model has been applied to nuclei which are proved to have octupole deformation, like the even-even Ra isotopes \cite{Rad8,Rad9}, as well as for Rn isotopes \cite{Rad10} whose negative parity states are interpreted as octupole vibrational states rather than states describing oscilations around a static octupole deformed shape.  The two sets of nuclei were described equally well, which leads to the conclusion that the proposed model is able to describe in an unified fashion negative parity spectra of pear shaped and octupole-spherical nuclei.

In the present paper we continue the project started in ref. \cite{Rad8} by extending the formalism to another two pairs of bands which are called conventionally, $\beta^{\pm}$ and $\gamma^{\pm}$. In fact, the phenomenological model called "coherent state model" (CSM), proposed for the description of the first three rotational bands of positive parity, ground, beta and gamma, is generalized by assuming for the intrinsic states associated to the ground, beta and gamma bands also an octupole deformation. From  each such a state one projects simultaneously the parity and angular momentum and consequently two bands of opposite parities are obtained. We want to see whether in the excited bands, $\beta^{\pm}, \gamma^{\pm}$
there are specific fingeprints of octupole deformation.

The project sketched above is achieved according to the following plan. In section 2, a brief review of the CSM model on which the formalism developed in the following sections hinges. The generalisation of the CSM formalism, by including the octupole degrees of freedom, is performed in Section 3.
Here, the projected states for the six bands are defined. The states involved 
form  an orthogonal set. In this restricted collective space, one defines an effective quadrupole and octupole boson Hamiltonian.
The electric transition probabilities between the eigenstates of the effective Hamiltonian are calculated in Section 4, by using very simple expressions for 
the E1, E2 and E3 transition operators. Numerical results for
$^{158}$Gd, $^{172}$Yb, $^{218}$Ra, $^{226}$Ra, $^{232}$Th, $^{238}$U, $^{238}$Pu are discussed in Section 5. The main conclusions are summarised in Section 6. 

\section{Brief review of the coherent states model for three interacting bands}
Two decades ago one of the present authors (A. A. R) proposed a phenomenological model to describe the main properties of the first three collective bands i.e., ground, beta and gamma bands \cite{Rad11,Rad12}.
The model space was generated through projection procedures from three orthogonal deformed states. The choice was made so that several criteria required by the existent data are fulfilled.
The states are built up with quadrupole bosons and therefore we are dealing
with those properties which are determined by the collective motion of the quadrupole degrees of freedom.

We suppose that the intrinsic ground state is described by  coherent states
of Glauber type corresponding to the zeroth component of the quadrupole boson
operator $b_{2\mu}$. The other two generating functions are the simplest polynomial excitations of the intrinsic ground state, chosen in such a way that the orthogonality condition is satisfied before and after projection.
To each intrinsic state one associates an infinite rotational band. In two of these bands the spin sequence is $0^+, 2^+, 4^+, 6^+,..$etc and therefore they correspond to the ground (the lowest one) and to the beta bands, respectively. The third one 
involves all angular momenta larger or equal to 2, and is describing, in the first order of approximation, the gamma band.
The intrinsic states depend on a real parameter d which simulates the nuclear deformation. In the spherical limit, i.e. d goes to zero, the projected states are multi-phonon states of highest, second and
third highest seniority, respectively. In the large deformation regime,
conventionally called rotational limit
(d equal to 3 means already a rotational limit), the model states behave like a Wigner function, which fully agrees the behavior prescribed by the liquid drop model. The correspondence between the states in the spherical and rotational limits is achieved by a smooth variation of the deformation parameter. This correspondence agrees perfectly with the semi-empirical rule of Sheline\cite{She}
 and Sakai\cite{Saka}
, concerning the linkage of the ground, beta and gamma band states and the member of multi-phonon states from the vibrational limit. This property is very important 
when one wants to describe the gross features of the reduced probabilities
for the intra and inter bands transitions.

In this restricted collective model space an effective boson Hamiltonian is constructed. A very simple Hamiltonian was found, which has only one off-diagonal matrix element, namely that one connecting the states from the ground and the gamma bands. 
\begin{eqnarray}
H_{CSM}&=&H_2^{\prime}+\lambda\hat{J}^2_2,\nonumber\\
H_2'&=&A_1(22\hat{N}_2+5\Omega^{\dagger}_{\beta'}\Omega_{\beta'})+A_2\Omega^{\dagger}_{\beta}\Omega_{\beta},
\end{eqnarray}
where $\hat{N}_2$ denotes the quadrupole boson number operator
\begin{equation}
\hat{N}_2=\sum_{-2\leq m\leq 2}b_{2m}^{\dagger}b_{2m}
\end{equation}
while $\Omega^{\dagger}_{\beta'}$ and  $\Omega^{\dagger}_{\beta}$ stand for the following second and third degree scalar polynomials:
 \begin{eqnarray}
  \Omega^{\dagger}_{\beta'}&=&(b^{\dagger}_2 b^{\dagger}_2)_0-\frac{d^2}{\sqrt{5}}\nonumber\\
  \Omega^{\dagger}_{\beta}&=&(b^{\dagger}_2 b^{\dagger}_2 b^{\dagger}_2)_0+\frac{3d}{\sqrt{14}}(b^{\dagger}_2 b^{\dagger}_2)_0-\frac{d^3}{\sqrt{70}}
  \end{eqnarray}
The angular momentum caried by the quadrupole bosons, is denoted by $\hat{J}_2$.
The boson states space is spanned by the projected states:
\begin{equation}
\phi^{(i)}_{JM}=N^{(i)}_JP^J_{MK}\psi_{i},~~i=g,\beta,\gamma,
\end{equation}
where the intrinsic states are:
\begin{equation}
\psi_g=e^{d(b^{\dag}_{20}-b_{20})}|0\rangle,~~
\psi_{\beta}=\Omega^{\dag}_{\beta}\psi_g,~~
\psi_{\gamma}=\Omega^{\dag}_{\gamma}\psi_g.
\end{equation}
where the excitation operator $\Omega^{\dag}_{\beta}$ was defined above, while the operator which excites the gamma band states is:
\begin{equation}
\Omega^{\dag}_{\gamma}=(b^{\dag}_2b^{\dag}_2)_{22}+d\sqrt{\frac{2}{7}}b^{\dag}_{22}.
\end{equation}
The eigenvalues of the effective Hamiltonian in the restricted space of projected states have been analytically studied in both spherical and rotational limit. Compact formulae for transition probabilities in the two extreme limits have been also derived. This model has been successfully applied for a large number of nuclei  from transitional and well deformed regions.
It is worth to mention that by varying the deformation parameter and the parameters defining the effective Hamiltonian one can realistically describe nuclei satisfying various symmetries like, SU(5) (Sm region)\cite{Rad13}, O(6) (Pt region)\cite{Rad11,Rad12}, SU3
(Th region)\cite{Rad6}, triaxial rotor (Ba, Xe isotopes)\cite{UliRad}.
This model has been extended by including the coupling to the individual degrees of freedom \cite{Rad14}. 
In this way the spectroscopic properties in the region of back-bending were quantitatively described.

Another extension of the CSM formalism is presented here and consists of
considering a composite system of quadrupole and octupole bosons. 
\section{Simultaneous description of positive and negative parity bands}
\label{sec:level2}
Here we generalize the CSM formalism by assuming that the intrinsic ground state exhibits not only a quadrupole deformation but also an octupole one. Since the other bands, beta and gamma, are excited from the ground state, they have also this property.
The octupole deformation is described by means of an axially symmetric coherent state for the octupole bosons $b^{\dag}_{30}$. Thus the intrinsic states for ground beta and gamma bands are:
\begin{equation}
\Psi_g=e^{f(b^{\dag}_{30}-b_{30})}e^{d(b^{\dag}_{20}-b_{20})}|0\rangle_{(3)}|0\rangle_{(2)},~
\Psi_{\beta}=\Omega^{\dag}_{\beta}\Psi_g,~
\Psi_{\gamma}=\Omega^{\dag}_{\gamma}\Psi_g. 
\end{equation}
The notation $|0\rangle_{(k)}$ stands for the vacuum state of the $2^k$-pole boson operators. Note that any of these states is a mixture of positive and negative parity states. Therefore they don't have good reflection symmetry.
Due to this feature the new extension of the CSM formalism has to project out not only the angular momentum but also the parity.
The parity projection affects only the factor function depending on octupole bosons.
For what follows it is convenient to treat this factor function separately.
The parity projected states are defined by:
\begin{equation}
\Psi^{(k)}=P^{(k)}e^{f(b^{\dag}_{30}-b_{30})}|0\rangle_{(3)},k=\pm
\end{equation}
where $P^{(k)}$ denotes the parity projection operator which is defined by its property that acting on a state consisting of a  series of boson operators
acting on the octupole vacuum, it selects only components with even powers in bosons if $k=+$ and odd components for $k=-$.
From the parity projected states one projects out, further, the components with good angular momentum:
\begin{equation}
\Psi^{(k)}_{J_3M_3}=N^{(k)}_{J_3}P^J_{M_30}\Psi^{(k)}.
\end{equation} 
The factor $N^{(k)}_{J_3}$ assures that the projected state has the norm equal to unity. Its expression is
\begin{equation}
\left(N^{(\pm)}_J\right)^{-2}=e^{-y_3}(2J+1){\cal I}^{(\pm)}_J(y_3),~y_3=f^2,
\end{equation}
where ${\cal I}^{(\pm)}$ stands  for the overlap functions:
\begin{eqnarray}
{\cal I}^{(+)}_J(y_3)&=&\int_0^1 P_J(x)ch\left[f^2P_3(x)\right]dx,\nonumber\\
{\cal I}^{(-)}_J(y_3)&=&\int_0^1 P_J(x)sh\left[f^2P_3(x)\right]dx.
\end{eqnarray}
with $P_J(x)$ denoting the Legendre polynomial of rank J.
The angular momentum projection operator is defined by:
\begin{equation}
P^J_{MK}=\frac{2J+1}{8\pi^2}\int D^{J^*}_{MK}(\Omega )\hat{R}(\Omega)d\Omega
\end{equation}
where the standard notations for the Wigner function and the rotation operator 
corresponding to the Eulerian angles $\Omega$, have been used.
The intrinsic states of good parity are obtained by applying the operator $P^{(k)}$ on the $\Psi_i, i=g,\beta,\gamma$.
\begin{equation}
\Psi^{(k)}_i=P^{(k)}\Psi_i,~i=g,\beta,\gamma,~k=\pm
\end{equation}
The member states of ground beta and gamma bands are projected from the corresponding intrinsic states.
\begin{eqnarray}
\varphi_{JM}^{(i,k)}&=&{\cal N}_J^{(i,k)}P_{MK_i}^J\Psi_i^{(k)},~~
K_i=2\delta_{i,\gamma},~~
k=\pm; i=g, \beta, \gamma,
\nonumber\\
&&J=(\delta_{i,g}+\delta_{i,\beta})(even \delta_{k,+}+ odd \delta_{k,-})
+\delta_{i,\gamma}(J\ge 2).
\end{eqnarray}
It can be shown that these projected states can be expressed by means of the octupole factor projected states and the projected states characterizing the CSM formalism.
\begin{equation}
\varphi^{(i,k)}_{JM}={\cal N}^{(i,k)}_J\sum_{J_2,J_3}\left(N^{(k)}_{J_3}
N^{(i)}_{J_2}\right)^{-1}C^{J_3~J_2~J}_{0~~0~~0}\left[\Psi^{(k)}_{J_3}
\varphi^{(i)}_{J_2}\right]_{JM}.
\end{equation}
The normalization factor has the expression:
\begin{equation}
({\cal N}_J^{(i,k)})^{-2}=\sum_{J_2,J_3}(N_{J_3}^{(k)}N_{J_2}^{(i)})^{-2}
(C_{0\hskip0.25cmK_i\hskip0.2cmK_i}^{J_3\hskip0.1cm J_2\hskip0.2cm J})^2,~
K_i=2\delta_{i,\gamma},~
k=\pm; i=g, \beta, \gamma.
\end{equation}
An effective boson Hamiltonian will be studied in the restricted collective space generated by the six sets of projected states.
Note that from each of the three intrinsic states, one generates by projection
 two sets of states, one of positive and one of negative parity.
When the octupole deformation goes to zero, the resulting states are just those characterizing the CSM model. In this limit we know already the effective quadrupole boson Hamiltonian. When the quadrupole deformation is going to zero
the system exhibits vibrations around an octupole deformed equilibrium shape.
We consider for the octupole Hamiltonian an harmonic structure since the non-harmonic octupole terms can be simulated by the quadrupole anharmonicities. As for the  coupling between quadrupole and octupole bosons, we suppose that this can be described by a product between the octupole boson number operator, $\hat {N}_3$, and the quadruople boson anharmonic terms which are involved in the CSM Hamiltonian. Also, two scalar terms depending on the angular momenta carried by the quadrupole ($J_2$) and 
($J_3$) octupole bosons, respectively are included. Thus the model Hamiltonian has the expression:
 
\begin{eqnarray}
H&=&H'_2+{\cal B}_1\hat{N}_3(22\hat{N}_2+5\Omega^{\dagger}_{\beta'}\Omega_{\beta'})+{\cal B}_2\hat{N}_3\Omega^{\dagger}_{\beta}\Omega_{\beta}
\nonumber\\
&&+{\cal B}_3\hat{N}_3+{\cal A}_{(J23)}\vec{J}_2\vec{J}_3+{\cal A}_J\vec{J}^2.
\end{eqnarray}
This Hamiltonian was used in Refs.\cite{Rad8,Rad9,Rad10} to study the ground and $K^{\pi}=0^-$
bands. Therein it was shown that the coupling term $\vec{J}_2\vec{J}_3$
is necessary in order to explain the low position of the state $1^-$ in the even-even Ra isotopes. Indeed, this term is attractive in the state $1^-$ while for other angular momenta it is repulsive.

The non-vanishing matrix elements of the model Hamiltonian between  projected states are given analytically in Appendix A.
Due to the specific structure of the CSM basis states the only
non-vanishing off-diagonal matrix elements are those connecting the states of 
the ground and gamma and of the $0^-$ and $2^-$ bands. Energies of these bands are obtained by diagonalizing 2x2 matrices while for the remaining states are given as expected values on the corresponding projected states.
\begin{eqnarray}
\left(\matrix{H^{(k)}_{gg;J}&H^{(k)}_{g\gamma;J}\cr
        H^{(k)}_{\gamma g;J}&H^{(k)}_{\gamma \gamma;J}}\right)
\left(\matrix{X^{(i,k)}_J\cr Y^{(i,k)}_J}\right)=\tilde{E}^{(i,k)}_J
\left(\matrix{X^{(i,k)}_J\cr Y^{(i,k)}_J}\right),i=g,\gamma; ~k=\pm
\end{eqnarray}
\begin{eqnarray}
\tilde {E}^{\beta,k}_J&=&\langle \varphi^{(\beta,k)}_J|H|\varphi^{(\beta,k)}_J\rangle ,k=\pm\nonumber\\
\tilde {E}^{\gamma,k}_J&=&\langle \varphi^{(\gamma,k)}_J|H|\varphi^{(\gamma,k)}_J\rangle , J=(odd)\delta_{k,+}+(even)\delta_{k,-}.
\end{eqnarray}
The matrix elements involved in the above equations are defined in Appendix A.
Energies depend on the structure coefficients ${\cal A}_k$,(k=1,2,J,(J23)) and 
${\cal B}_k$
,(k=1,2,3) defining the model Hamiltonian and the two deformation parameters, d and f. In the applications we consider here, 
the parameter ${\cal B}_2$ is not considered in the fitting procedure since data for the $\beta^-$ band are missing.
Therefore there  are 8 parameters which will be determined in Section 5, by  fitting the data for excitation energies with the theoretical energies normalized to the ground state energy:
\begin{equation}
E^{(i,k)}_J=\tilde{E}^{(i,k)}_J-\tilde{E}^{(g,+)}_0,i=g,\beta,\gamma;~k=\pm
\end{equation}
The final states modeling the member states of ground and gamma bands of positive and negative parity states are:
\begin{eqnarray}
\Psi^{(g,k)}_{JM}&=&X^{(g,k)}_J\varphi^{(g,k)}_{JM}+Y^{(g,k)}_J\varphi^{(\gamma,k)}_{JM},J=\rm{even}\delta_{k,+}+\rm{odd}\delta_{k,-}\nonumber\\
\Psi^{(\gamma,k)}_{JM}&=&X^{(\gamma,k)}_J\varphi^{(g,k)}_{JM}+Y^{(\gamma,k)}_J\varphi^{(\gamma,k)}_{JM}.
\end{eqnarray}
These states together with $\varphi^{(\beta,k)}_{JM}$ and $\varphi^{(\gamma,+)}
_{JM}$ (J=odd), $\varphi^{(\gamma,-)}_{JM}$ (J=even)  are used to calculate, in the next section, the reduced probabilities for the E1, E2 and E3 transitions.
\section{Electric multipole transitions}
The states defined in the previous section are connected through electric type  transitions. Here we are interested in describing the intra band E2 transitions as well as the E1 and E3 transitions connecting the positive and negative parity states generated from the same intrinsic deformed wave function.
In order to outline the main virtues of the model states we consider the transition operators in the lowest order in  boson operators.
Thus the E2 and E3 transition operators are linear in the quadrupole and octupole boson operators, respectively:
\begin{equation}
Q_{\lambda\mu}=q_{\lambda}(b^{\dag}_{\lambda\mu}+(-)^\mu b_{\lambda,-\mu}),
~\lambda=2,3.
\end{equation}
The lowest order for the E1 transition operator is
\footnote{throughout this paper the Rose \cite{Rose} conventions for reduced matrix elements are used}:
\begin{equation}
T_{1\mu}=\sum_{\mu_2,\mu_3}{C^{3~2~1}_{\mu_3~\mu_2~\mu}(b^{\dag}_{3\mu_3}+(-)^\mu b_{3,-\mu_3}) (b^{\dag}_{2\mu_2}+(-)^\mu b_{2,-\mu_2})}
\end{equation}
The reduced probability for the E1 transitions
$I^-_i\rightarrow J^+_i,(i=g,\beta,\gamma)$ is readily obtained by squaring the corresponding reduced matrix element:
\begin{eqnarray}
B(E1;I^-_i\rightarrow J^+_i)&=&\left[\langle \Psi^{(i,-)}_I||T_1||\Psi^{(i,+)}_J\rangle \right]^2,i=g,\gamma,~I=\rm{odd},~~J=\rm{even} \nonumber\\
B(E1;I^-_i\rightarrow J^+_i)&=&\left[\langle \varphi^{(i,-)}_I||T_1||\varphi^{(i,+)}_J\rangle \right]^2, i=\beta 
~\rm{and}~ i=\gamma~~\rm{ for}~~ I=\rm{even}~~\rm{ and}~~ J=\rm{odd},\nonumber\\
B(E1;I^-_i\rightarrow J^+_i)&=&\left[\langle \varphi^{(i,-)}_I||T_1||\Psi^{(i,+)}_J\rangle \right]^2, i=\gamma~ \rm{for}~~ I=\rm{even}~~ and~~ J=\rm{even}\nonumber\\
B(E1;I^-_i\rightarrow J^+_i)&=&\left[\langle \Psi^{(i,-)}_I||T_1||\varphi^{(i,+)}_J\rangle \right]^2, i=\gamma~~ \rm{for}~~ I=\rm{odd}~~\rm{ and}~~ J=\rm{even}.
\end{eqnarray}
The reduced probabilities for the quadrupole transitions between states belonging to the same band are given by:
\begin{eqnarray}
B(E2;J^{k}_i\rightarrow {J^{\prime}}^{k}_i)&=&\left[\Psi^{(i,k)}_J||Q_2||
\Psi^{(i,k)}_{J^{\prime}}\right]^2,i=g,\gamma, J,J'=\rm{even}\delta_{k,+}
+\rm{odd}\delta_{k,-}\\
B(E2;J^{k}_i\rightarrow {J^{\prime}}^{k}_i)&=&\left[\varphi^{(i,k)}_J||Q_2||
\varphi^{(i,k)}_{J^{\prime}}\right]^2,i=g,\gamma ( J,J'=\rm{odd}\delta_{k,+}
+\rm{even}\delta_{k,-}),\beta\nonumber\\
B(E2;J^{k}_i\rightarrow {J^{\prime}}^{k}_i)&=&\left[\varphi^{(i,k)}_J||Q_2||
\Psi^{(i,k)}_{J^{\prime}}\right]^2,i=g,\gamma( J=\rm{odd}\delta_{k,+}
+\rm{even}\delta_{k,-},J'=\rm{odd}\delta_{k,-}+\rm{even}\delta_{k,+})\nonumber\\
B(E2;J^{k}_i\rightarrow {J^{\prime}}^{k}_i)&=&\left[\Psi^{(i,k)}_J||Q_2||
\varphi^{(i,k)}_{J^{\prime}}\right]^2,i=g,\gamma( J'=\rm{odd}\delta_{k,+}
+\rm{even}\delta_{k,-},J=\rm{even}\delta_{k,+}+\rm{odd}\delta_{k,-})\nonumber
\end{eqnarray}
States from partner bands can also be connected through the octupole transition operator. The corresponding B(E3)  values are defined as follows:
\begin{eqnarray}
B(E3;J^{(+)}_{\beta}\rightarrow {J^{\prime}}^{-}_\beta)&=&
\left[\langle \varphi^{(\beta,+)}_J||Q_3||\varphi^{(\beta,-)}_{J'}\rangle\right]^2
\nonumber\\
B(E3;J^{(+)}_{i}\rightarrow {J^{\prime}}^{-}_i)&=&
\left[\langle \Psi^{(i,+)}_J||Q_3||\Psi^{(i,-)}_{J'}\rangle\right]^2,
i=g,\gamma (J=\rm{even},~J^{\prime}=\rm{odd}),\nonumber\\
B(E3;J^{(+)}_{\gamma}\rightarrow {J^{\prime}}^{-}_\gamma)&=&
\left[\langle \varphi^{(\gamma,+)}_J||Q_3||\varphi^{(\gamma,-)}_{J'}\rangle\right]^2,
J=\rm{odd},~J^{\prime}=\rm{even}\nonumber\\
B(E3;J^{(+)}_{\gamma}\rightarrow {J^{\prime}}^{-}_\gamma)&=&
\left[\langle \Psi^{(\gamma,+)}_J||Q_3||\varphi^{(\gamma,-)}_{J'}\rangle\right]^2,
J,J^{\prime}=\rm{even},\nonumber\\
B(E3;J^{(+)}_{\gamma}\rightarrow {J^{\prime}}^{-}_\gamma)&=&
\left[\langle \varphi^{(\gamma,+)}_J||Q_3||\Psi^{(\gamma,-)}_{J'}\rangle\right]^2,
J,J^{\prime}=\rm{odd}.
\end{eqnarray}

\section{Numerical results}
\label{sec:level5}
The formalism described in the previous sections has been applied to seven even-even nuclei for which there are data concerning the bands under consideration.
These nuclei are:$^{158}$Gd, $^{172}$Yb, $^{218}$Ra, $^{226}$Ra, $^{232}$Th, $^{238}$U, $^{238}$Pu. Among these nuclei there are two, $^{218}$Ra, $^{226}$Ra, which are known to have octupole deformation. The negative parity states in the remaining nuclei have a vibrational character. Including them in the present study, has the goal to prove the capacity of this model to describe  the octupole deformation in the vanishing limit which results in providing an unified picture for spherical and octupole deformed nuclei. Before starting the quantitative analysis of our results, we would like to study the structure of the projected states, i.e, to see, for a given angular momentum, what is the average value for the quadrupole and octupole boson number operators. This allows us to draw definite conclusions concerning the partial contribution of various terms of the model Hamiltonian to the state energies. Since the ground bands have been already studied in Ref. \cite{Rad8} we omit them in the present study. Thus the average values for the $\hat{N}_2$ and $\hat{N}_3$ operators
on the states of gamma and beta bands are plotted as functions of the octupole deformation for a small ( or a large) value of the quadrupole deformation, in Figs.1-3.  

We also studied these average values as function of d for small (f=0.2) and large (f=2.) octupole deformations. However, for the sake of saving space we don't give them here in detail, but describe shortly the results.
For small values of octupole deformation (f=0.2) the spectrum of $\hat{N}_2$ exhibits several interesting features. In the region of small quadrupole deformation the levels have high degeneracy due to competing contributions of both quadrupole and octupole boson operators. For large quadrupole deformations the normalized average values have a typical rotational behavior.
In negative beta band the state $1^-$ is higher in energy than the state $3^-$
for low values of d, while for $d>2$  a normal ordering is shown up.
Note that for low d the states $1^-,5^-$ are degenerate. This reflects the quadrupole octupole two phonon structure of these two states. 
 For $0.5\le d\le 2.5$ the gamma band shows a doublet structure $(3^+,4^+);(5^+,6^+);(7^+,8^+)$, etc. Increasing further  the quadrupole deformation parameter the staggering disappears while for d beyond 3 a different doublet structures appears $(2^+,3^+);(4^+,5^+);(6^+,7^+)$, etc. These properties are reminiscent from the CSM model.
For large values of d and f=0.2 the averaged $\hat{N}_3$ is 0 for positive parity gamma  and beta  and 1 for negative gamma  and beta  states.
For f=2 and small d the averaged $\hat{N}_2$ does not depend too much on angular momentum while for large d a rotational spectrum is obtained. By contrast the averaged $\hat{N}_3$ has a large split with angular momentum for small d and a compressed spectrum for large d.

 The dependence of the averaged values for $\hat{N}_2,\hat{N}_3$ on octupole deformation is shown in Fig. 1. The average number of quadrupole bosons is independent on angular momentum for f larger than 2 in  beta band. Indeed the average value is 3 for any angular momentum. The behaviour of the average value of $\hat{N}_2$ in gamma band is similar with that in beta band with the difference that the constant value is 2 instead of 3. Because of that, here we don't show the corresponding plot 
The average of  $\hat{N}_3$ exhibits a nice shell structure in both beta and gamma bands. In order to see that more clearly we represented, in Fig. 4, the sequence of these average values for d=0.2 and f=1.4. The shell structure is more pronounced in gamma band where the shells of positive and negative parities appear in a sequential order.  We note that in each upper shells there is one intruder state. 
Complementary features for octupole and quadrupole degrees of freedom are shown in Fig. 2) for large d(=2.). Indeed, while $\hat{N}_3$ is degenerate with respect to angular momentum for small f (Fig. 2 c), d) and Figs. 3 a), b)), in the region of large f a rotational behavior
is set on. In contradistinction to $\hat{N}_3$, $\hat{N}_2$ has a quasi-rotational spectrum for small f while for large f the split over J is quite small
 (see Figs. 2 a), b), e), f)).
Another term involved in the model Hamiltonian is the product 
$\hat{N}_2\hat{N}_3$. Its average values are considered for small octupole deformations (f=0.2) in Figs. 3 c), d), e), f). From there one draws the conclusion that this term does not contribute at all to the expected energy of positive parity beta and gamma states, for d larger that 1.75 while for negative parity states the contribution increases with d for the interval mentioned above.
The detailed analysis mentioned  above is very important when one searches for the optimal parameters. 
The parameters involved in the excitation energies were determined by a least square fit. The results are listed in Table 1. 

\newpage
Table 1.  
{\small The parameters defining the model Hamiltonian yielded by the fitting procedure are given in units of KeV. The deformation parameters \emph{d} and \emph{f} 
are adimensional.} 
\begin{center}
\begin{tabular}{ p{1cm} rrrrrrr}\hline
&$^{158}Gd$   &$\hskip0.2cm$$^{172}Yb$       &$\hskip0.2cm$$^{218}Ra$     &
\hskip0.2cm$^{226}Ra$     &$^{232}Th$     &$^{238}U$      &$^{238}Pu$\\ \hline        
              
$d$                     &3.00          &3.68      &1.40       &3.15     &3.25           &3.90           &3.90\\      
$f$                     &0.30         &0.30      &0.30       &0.65      &0.30           &0.60           &0.30\\
${\cal A}_1$         &21.49        &26.94     &16.86    &21.07       &14.26          &20.64          &18.84 \\
${\cal A}_2$         &-12.29       &-17.68    &-23.43   &-17.41      &-8.33          &-9.72          &-8.63\\
${\cal A}_J$         &3.10          &4.72      &1.81     &0.51       &2.26           &1.55           &2.23\\   
${\cal A}_{(J23)}$   &34.70         &38.49     &18.09    &9.53       &11.93          &20.45          &10.77\\
${\cal B}_1$         &-11.84       &-24.25    &-7.13   &-1.95        &-6.17          &-10.77         &-8.39\\      
${\cal B}_3$         &3 673.27     &8 711.1   &786.24    &641.08     &2 093.34
&4 239.42        &3 308.60\\ \hline 
\end{tabular}    
\end{center}
For illustration we selected two cases, $^{226}$Ra and $^{232}$Th, for which the predicted energies are compared with the corresponding experimental data in a plot, Figs. 5, 6, showing them as function of the angular momentum. Energies for these nuclei are produced with octupole deformation which are quite different, namely 0.65 for $^{226}$Ra and 0.3 for $^{232}$Th. The three pairs of bands in 
$^{226}$Ra have  the specific feature that the doublet partners, starting from a certain angular momentum, have similar moments of inertia. Indeed, their energies are interleaved and lie on a curve which is close to a straight line. The parity doublet structure is more pronounced in beta bands than in ground bands.
In gamma bands we have a typical case of parity doublets. Indeed, for $J\ge8$ the states of equal angular momenta but different parities are staggered together in a doublet structure.
As for $^{232}$Th, it has a larger quadrupole deformation than $^{226}$Ra but smaller octupole deformation. It is worth mentioning that for this nucleus we found 55 experimental excitation energies which were described by the predictions
of our model with a high accuracy. The energies for the two ground bands lie on different curves which intersect each other at high angular momentum 
(about $29^-$). Beta bands reach each other much earlier, around $J=12$, and go together
up to $J=20$ and then go apart from each other. The gamma bands get similar moments of inertia starting with $20^+$.
The presence of octupole deformation is usually judged from the plot of the energy displacement function:
\begin{equation}
\delta E(J^-)=E(J^-)-\frac{(J+1)E((J-1)^+)+JE((J+1)^+)}{2J+1}
\end{equation}
where
$E(J^{\pi})$ is the energy of the state of angular momentum J and parity $\pi$.
If the partner bands, in a certain interval of angular momentum, have a $J(J+1)$
behavior and moreover they have similar moments of inertia, the displacement function vanishes in the given interval.
This function is plotted  for $^{226}$Ra and $^{232}$Th in Figs. 7 and 8, respectively.
From Fig. 10 one sees that the ground bands show an excellent octupole deformed structure in the interval $11^--27^-$. The common feature of the two cases shown in Fig. 10 and Fig. 11 is that the octupole deformation shows up first in beta band then in gamma band and latest in ground band.
The predicted energies for the six bands are compared with the available data in several tables (Tables 2-7). By inspecting these tables one may conclude upon a very good agreement between predicted energies and the corresponding data.
In Ref.\cite{Fern} the state $2^+$ from the ground band is identified at 0.352
MeV which is very close to our prediction for the excitation energy of this 
state.
While in the figures commented above we notice that partner bands reach each other at a certain angular momentum,
the results collected in tables show some new features in excited bands. Indeed, there are three cases, 
$^{238}$U,$^{238}$Pu and $^{172}$Yb where the band $\beta^-$ is lower in energy than the $\beta^+$ one. This is opposite to what  have seen for the ground bands where the positive parity band  is lower than the partner negative parity band. However before having data for the beta minus band energies it is hard to draw definite conclusion on this matter, given the fact that energies in this band are free of any adjustable parameter. Another peculiar aspect appears in $^{238}$U where the $\beta^-$ band intersects the $K^{\pi}=0^-$-ground band at $J^{\pi}=11^-$. Moreover by glacing at Table 5 one sees that the data for the levels $1^-,3^-,5^-$ are very close to the predictions for corresponding levels in $\beta^-$ band. Therefore 
in order to decide to what band one should attached the three observed states,
data about decaying properties of these states are necessary.

It is worth spending few words in connection with the case of $^{172}$Yb. Indeed this nucleus seems not to have static octupole deformation in its ground state. Therefore the ground band $0^-$ is highly excited. This corresponds to a potential energy in octupole deformation variable, which has a deep minimum for 
vanishing octupole deformation. This reflects in the high value of the 
coefficient ${\cal B}_3$ which, according to Figs. 1, 2, results in shifting
the negative bands with a large quantity.

Table 2. - 
{\small Experimental (left column) and predicted (right column) energies for the ground band. The energy values are given
in MeV. Data are from}:\cite{Sug,Blo,Lee} ($^{158}$Gd),\cite{Gre,Bur,Bal} ($^{172}$Yb), \cite{Gai1,Gono} ($^{218}$Ra), \cite{Coc1,Wol,Ell} ($^{226}$Ra),  
\cite{Coc1,Sim,Gro}($^{232}$Th), \cite{Ale,Shu} ($^{238}$U), \cite{Gai1,Sim,Shu} ($^{238}$Pu).
\begin{flushleft}
\begin{tabular}{ccccccccccccccc}\hline
$J^{\pi}$&\multicolumn{2}{c}{$^{158}Gd$}&\multicolumn{2}{c}{$^{172}Yb$}&\multicolumn{2}{c}{$^{218}Ra$}&\multicolumn{2}{c}{$^{226}Ra$}&
\multicolumn{2}{c}{$^{232}Th$}&\multicolumn{2}{c}{$^{238}U$}&\multicolumn{2}{c}
{$^{238}Pu$}\\ \hline        
$2^+$   &0.080&0.084    &0.079&0.078    &0.389&0.347    &0.068&0.060    &0.049&0.049    &0.045&0.043    &0.044&0.044\\  
$4^+$   &0.261&0.274    &0.260&0.260    &0.741&0.720    &0.212&0.193    &0.162&0.162    &0.148&0.144    &0.146&0.146\\
$6^+$   &0.540&0.555    &0.540&0.540    &1.122&1.117    &0.417&0.391    &0.333&0.332    &0.307&0.300    &0.303&0.303\\
$8^+$   &0.904&0.913    &0.912&0.912    &1.547&1.535    &0.670&0.642    &0.557&0.553    &0.518&0.506    &0.513&0.514\\
$10^+$  &1.351&1.337    &1.370&1.370    &1.962&1.965    &0.961&0.935    &0.827&0.819    &0.776&0.761    &0.773&0.774\\
$12^+$  &1.867&1.818    &1.907&1.906    &2.392&2.404    &1.282&1.264    &1.137&1.125    &1.077&1.058    &1.079&1.079\\
$14^+$  &-&2.348        &2.518&2.516    &2.827&2.845    &1.629&1.620    &1.483&1.467    &1.415&1.395    &1.427&1.427\\
$16^+$  &-&2.923        &3.198&3.194    &3.286&3.285    &1.999&2.000    &1.859&1.841    &1.788&1.767    &1.816&1.814\\
$18^+$  &-&3.538        &-&3.934        &-&3.724        &2.390&2.400    &2.263&2.244    &2.191&2.170    &2.241&2.237\\
$20^+$  &-&4.191        &-&4.734        &-&4.160        &2.801&2.817    &2.692&2.675    &2.619&2.602    &-&2.693\\
$22^+$  &-&4.879        &-&5.589        &-&4.595        &3.233&3.249    &3.144&3.132    &3.067&3.060    &-&3.182\\
$24^+$  &-&5.600        &-&6.496        &-&5.030        &3.686&3.695    &3.620&3.612    &3.535&3.539    &-&3.699\\
$26^+$  &-&6.354        &-&7.453        &-&5.467        &4.159&4.153    &4.116&4.116    &4.017&4.037    &-&4.244\\
$28^+$  &-&7.139        &-&8.459        &-&5.906        &4.651&4.621    &4.632&4.641    &4.517&4.550    &-&4.816\\
$30^+$  &-&7.954        &-&9.509        &-&6.347        &-&5.089        &5.162&5.187    &5.034&5.065    &-&5.411\\ \hline

\end{tabular}
\end{flushleft}

\newpage
 On the other hand the coupling term 
whose strength is ${\cal B}_1$ is also large and attractive which means that this term decreases the energy shift mentioned above. The competition between
the two opposite effects is decided by the position of the ground band of negative parity. Since the effect of the coupling term is very large in $\beta^-$
band, this is strongly pressed down. This explains the relative position of
the two $K^{\pi}=0^-$ bands. This is the only case where $\beta^-$ is lower than the negative parity band which is partner of the ground band.

Table 3. -
{\small The same as in table 2 but for the $K^{\pi}=0^-$ band}.
\begin{flushleft}
\begin{tabular}{ccccccccccccccc}\hline
$J^{\pi}$&\multicolumn{2}{c}{$^{158}Gd$}&\multicolumn{2}{c}{$^{172}Yb$}&\multicolumn{2}{c}{$^{218}Ra$}&\multicolumn{2}{c}{$^{226}Ra$}&
\multicolumn{2}{c}{$^{232}Th$}&\multicolumn{2}{c}{$^{238}U$}&\multicolumn{2}{c}
{$^{238}Pu$}\\ \hline        
$1^-$&1.264&1.267       &1.600&1.599    &(0.713)&0.726    &0.254&0.243    &0.714&0.709    &0.681&0.637    &0.605&0.604\\
$3^-$&1.403&1.402       &1.711&1.711    &0.794&0.750    &0.322&0.312    &0.774&0.773    &0.732&0.700    &0.661&0.661\\
$5^-$&1.639&1.635       &-&1.910        &1.038&1.026    &0.447&0.438    &0.884&0.888    &0.827&0.811    &0.763&0.763\\
$7^-$&-&1.954           &-&2.187        &1.341&1.363    &0.627&0.620    &1.043&1.050    &0.966&0.969    &-&0.908\\
$9^-$&-&2.342           &-&2.537        &1.695&1.730    &0.858&0.854    &1.249&1.256    &1.151&1.169    &-&1.095\\
$11^-$&-&2.790          &-&2.951        &2.110&2.119    &1.133&1.134    &1.499&1.502    &1.378&1.409    &-&1.321\\
$13^-$&-&3.286          &-&3.423        &2.527&2.525    &1.448&1.454    &1.785&1.782    &1.649&1.684    &-&1.585\\
$15^-$&-&3.825          &-&3.948        &2.967&2.945    &1.796&1.807    &2.102&2.095    &1.959&1.992    &-&1.884\\
$17^-$&-&4.402          &-&4.523        &3.390&3.378    &2.175&2.187    &2.445&2.437    &2.306&2.330    &-&2.216\\
$19^-$&-&5.014          &-&5.143        &3.797&3.819        &2.579&2.591    &2.813&2.806    &2.687&2.694    &-&2.578\\
$21^-$&-&5.657          &-&5.806        &4.260&4.267        &3.006&3.015    &3.204&3.200    &3.104&3.082    &-&2.970\\
$23^-$&-&6.332          &-&6.512        &-&4.719        &3.455&3.455    &3.616&3.618    &3.548&3.493    &-&3.389\\
$25^-$&-&7.035          &-&7.258        &-&5.173        &3.922&3.910    &4.051&4.059    &-&3.923        &-&3.834\\
$27^-$&-&7.767          &-&8.044        &-&5.629        &4.406&4.377    &-&4.521        &-&4.372        &-&4.304\\
$29^-$&-&8.490          &-&8.723        &-&6.085        &-&4.780        &-&4.947        &-&4.630        &-&4.570\\ \hline

\end{tabular}
\end{flushleft}

\newpage

Table 4. 
{\small The same as in table 2 but for the $\beta^+$ band ($K^{\pi}=0^+$)} .
\begin{flushleft}
\begin{tabular}{ccccccccccccccc}\hline
$J^{\pi}$&\multicolumn{2}{c}{$^{158}Gd$}&\multicolumn{2}{c}{$^{172}Yb$}&\multicolumn{2}{c}{$^{218}Ra$}&\multicolumn{2}{c}{$^{226}Ra$}&
\multicolumn{2}{c}{$^{232}Th$}&\multicolumn{2}{c}{$^{238}U$}&\multicolumn{2}{c}
{$^{238}Pu$}\\ \hline  
$0^+$&1.196&1.196       &1.043&1.043    &0.974&0.974    &0.825&0.825    &0.731&0.731            &0.925&0.925    &0.942&0.942\\
$2^+$&1.260&1.266       &1.118&1.113    &-&1.189        &0.874&0.871    &0.774&0.774            &0.967&0.963    &0.983&0.982\\
$4^+$&1.407&1.428       &1.288&1.276    &-&1.473        &-&0.977        &0.873&0.872            &1.057&1.051    &-&1.075\\
$6^+$&-&1.674           &1.538&1.528    &-&1.795        &-&1.137        &1.023&1.023            &-&1.187        &-&1.221\\
$8^+$&-&1.994           &1.854&1.866    &-&2.136        &-&1.343        &1.222&1.222            &-&1.370        &-&1.416\\
$10^+$&-&2.379          &2.213&2.286    &-&2.486        &-&1.591        &1.469&1.465            &-&1.595        &-&1.658\\
$12^+$&-&2.823          &2.607&2.782    &-&2.834        &-&1.871        &1.755&1.749            &-&1.860        &-&1.945\\
$14^+$&-&3.320          &3.304&3.350    &-&3.173        &-&2.181        &2.081&2.069            &-&2.162        &-&2.273\\
$16^+$&-&3.862          &-&3.985        &-&3.498        &-&2.514        &2.441&2.422            &-&2.497        &-&2.641\\
$18^+$&-&4.448          &-&4.683        &-&3.808        &-&2.867        &2.832&2.806            &-&2.861        &-&3.046\\
$20^+$&-&5.073          &-&5.441        &-&4.105        &-&3.238        &3.249&3.219            &-&3.252        &-&3.485\\
$22^+$&-&5.734          &-&6.256        &-&4.393        &-&3.625        &-&3.659                &-&3.665        &-&3.956\\
$24^+$&-&6.430          &-&7.124        &-&4.676        &-&4.026        &-&4.123                &-&4.100        &-&4.458\\
$26^+$&-&7.159          &-&8.042        &-&4.955        &-&4.439        &-&4.612                &-&4.554        &-&4.989\\
$28^+$&-&7.919          &-&9.010        &-&5.233        &-&4.865        &-&5.123                &-&5.026        &-&5.548\\
$30^+$&-&8.709          &-&10.026       &-&5.510        &-&5.304        &-&5.656                &-&5.520        &-&6.133\\ \hline

\end{tabular}
\end{flushleft}

\newpage
Table 5.
{\small Predicted energies for the $\beta^-$ band ($K^{\pi}=0^-$)} .
\begin{flushleft}
\begin{tabular}{p{1cm}p{2cm}p{2cm}p{2cm}p{2cm}p{2cm}p{2cm}p{2cm}}\hline
$J^{\pi}$       &$^{158}Gd$     &$^{172}Yb$     &$^{218}Ra$     &$^{226}Ra$ &$^{232}Th$  &$^{238}U$  &$^{238}Pu$\\  \hline  
$1^-$&1.455     &0.557  &1.033  &0.892  &0.910  &0.682  &0.816\\
$3^-$&1.580     &0.661  &1.120  &0.952  &0.970  &0.740  &0.870\\
$5^-$&1.795     &0.844  &1.330  &1.061  &1.076  &0.841  &0.965\\
$7^-$&2.089     &1.101  &1.594  &1.218  &1.226  &0.986  &1.101\\
$9^-$&2.451     &1.424  &1.891  &1.419  &1.416  &1.170  &1.276\\
$11^-$&2.870    &1.809  &2.209  &1.662  &1.644  &1.391  &1.489\\
$13^-$&3.337    &2.249  &2.541  &1.942  &1.906  &1.646  &1.737\\
$15^-$&3.846    &2.740  &2.884  &2.253  &2.200  &1.930  &2.020\\
$17^-$&4.393    &3.276  &3.233  &2.592  &2.523  &2.243  &2.335\\
$19^-$&4.975    &3.857  &3.583  &2.954  &2.872  &2.581  &2.680\\
$21^-$&5.589    &4.478  &3.933  &3.337  &3.248  &2.942  &3.054\\
$23^-$&6.234    &5.140  &4.277  &3.737  &3.647  &3.323  &3.455\\
$25^-$&6.908    &5.841  &4.614  &4.152  &4.069  &3.724  &3.883\\
$27^-$&7.611    &6.580  &4.943  &4.580  &4.513  &4.142  &4.336\\
$29^-$&8.393    &7.514  &5.263  &5.027  &5.002  &4.664  &4.854\\ \hline

\end{tabular}
\end{flushleft}

\newpage

Table 6.
{\small The same as in table 2 but for the $\gamma^+$ band ($K^{\pi}=2^+$)}.
\begin{flushleft}
\begin{tabular}{ccccccccccccccc}\hline
$J^{\pi}$&\multicolumn{2}{c}{$^{158}Gd$}&\multicolumn{2}{c}{$^{172}Yb$}&\multicolumn{2}{c}{$^{218}Ra$}&\multicolumn{2}{c}{$^{226}Ra$}&
\multicolumn{2}{c}{$^{232}Th$}&\multicolumn{2}{c}{$^{238}U$}&\multicolumn{2}{c}
{$^{238}Pu$}\\ \hline  
$2^+$   &1.187&1.195    &1.465&1.481    &1.072&0.992    &1.156&1.142            &0.785&0.790    &1.061&1.069    &1.029&1.029\\
$3^+$   &1.266&1.265    &1.549&1.553    &-&1.162        &-&1.190                &0.829&0.833    &1.106&1.108    &1.069&1.070\\
$4^+$   &1.358&1.358    &1.657&1.648    &-&1.335        &-&1.255                &0.891&0.890    &1.168&1.160    &1.126&1.123\\
$5^+$   &1.482&1.472    &1.780&1.766    &-&1.522        &-&1.333                &0.961&0.959    &-&1.224        &-&1.190\\
$6^+$   &-&1.607        &-&1.908        &-&1.710        &-&1.425                &1.049&1.043    &-&1.300        &-&1.270\\
$7^+$   &-&1.758        &-&2.070        &-&1.911        &-&1.529                &1.142&1.137    &-&1.387        &-&1.362\\
$8^+$   &-&1.932        &-&2.255        &-&2.107        &-&1.646                &1.259&1.246    &-&1.487        &-&1.466\\
$9^+$   &-&2.116        &-&2.459        &-&2.318        &-&1.771                &1.371&1.362    &-&1.597        &-&1.582\\
$10^+$  &-&2.324        &-&2.686        &-&2.519        &-&1.911                &1.512&1.494    &-&1.719        &-&1.711\\
$11^+$  &-&2.536        &-&2.928        &-&2.736        &-&2.053                &1.641&1.630    &-&1.849        &-&1.848\\
$12^+$  &-&2.776        &-&3.196        &-&2.938        &-&2.213                &1.801&1.783    &-&1.993        &-&2.000\\
$13^+$  &-&3.012        &-&3.473        &-&3.156        &-&2.368                &-&1.936    &-&2.141        &-&2.158\\
$14^+$  &-&3.281        &-&3.779        &-&3.357        &-&2.545                &2.117&2.109    &-&2.304        &-&2.332\\
$15^+$  &-&3.538        &-&4.089        &-&3.574        &-&2.710                &-&2.278    &-&2.469        &-&2.508\\
$16^+$  &-&3.834        &-&4.432        &-&3.773        &-&2.903                &2.446&2.469    &-&2.650        &-&2.703\\
$17^+$  &-&4.109        &-&4.771        &-&3.986        &-&3.077                &-&2.651    &-&2.829        &-&2.896\\
$18^+$  &-&4.431        &-&5.149        &-&4.184        &-&3.283                &2.767&2.861    &-&3.028        &-&3.111\\
$19^+$  &-&4.721        &-&5.515        &-&4.391        &-&3.463                &-&3.055        &-&3.219        &-&3.320\\
$20^+$  &-&5.067        &-&5.927        &-&4.588        &-&3.683                &-&3.281        &-&3.434        &-&3.554\\ \hline 

\end{tabular}
\end{flushleft}

\newpage
Table 7.
{\small The same as in table 2 but for the $\gamma^-$ band ($K^{\pi}=2^-$)}.
\begin{flushleft}
\begin{tabular}{ccccccccccccccc}\hline
$J^{\pi}$&\multicolumn{2}{c}{$^{158}Gd$}&\multicolumn{2}{c}{$^{172}Yb$}&\multicolumn{2}{c}{$^{218}Ra$}&\multicolumn{2}{c}{$^{226}Ra$}&
\multicolumn{2}{c}{$^{232}Th$}&\multicolumn{2}{c}{$^{238}U$}&\multicolumn{2}{c}
{$^{238}Pu$}\\ \hline  
$2^-$&1.794&1.786       &1.757&1.757    &-&1.162        &1.239&1.253    &-&1.145    &1.129&1.148    &-&1.168\\
$3^-$&1.861&1.861       &1.822&1.822    &-&1.209        &-&1.291        &1.183&1.182    &1.171&1.184    &1.202&1.201\\
$4^-$&1.954&1.960       &-&1.907        &-&1.292        &-&1.342        &-&1.230    &1.233&1.231    &-&1.244\\
$5^-$&-&2.081           &-&2.013        &-&1.412        &-&1.405        &1.294&1.291    &1.286&1.289    &-&1.298\\
$6^-$&-&2.222           &-&2.138        &-&1.556        &-&1.481        &-&1.362        &1.382&1.358    &-&1.362\\
$7^-$&-&2.383           &-&2.282        &-&1.717        &-&1.570        &-&1.445        &-&1.438        &-&1.437\\
$8^-$&-&2.561           &-&2.444        &-&1.888        &-&1.670        &-&1.538        &-&1.528        &-&1.522\\
$9^-$&-&2.756           &-&2.623        &-&2.069        &-&1.782        &-&1.641        &-&1.629        &-&1.616\\
$10^-$&-&2.964          &-&2.818        &-&2.255        &-&1.904        &-&1.753        &-&1.738        &-&1.720\\
$11^-$&-&3.188          &-&3.029        &-&2.447        &-&2.040        &-&1.876        &-&1.857        &-&1.835\\
$12^-$&-&3.423          &-&3.255        &-&2.645        &-&2.181        &-&2.006        &-&1.985        &-&1.957\\
$13^-$&-&3.673          &-&3.495        &-&2.845        &-&2.337        &-&2.148        &-&2.121        &-&2.090\\
$14^-$&-&3.929          &-&3.748        &-&3.050        &-&2.493        &-&2.294        &-&2.265        &-&2.229\\
$15^-$&-&4.202          &-&4.015        &-&3.258        &-&2.668        &-&2.452        &-&2.418        &-&2.380\\
$16^-$&-&4.477          &-&4.293        &-&3.469        &-&2.838        &-&2.612        &-&2.575        &-&2.535\\
$17^-$&-&4.771          &-&4.585        &-&3.680        &-&3.030        &-&2.786        &-&2.744        &-&2.703\\
$18^-$&-&5.063          &-&4.887        &-&3.896        &-&3.210        &-&2.959        &-&2.914        &-&2.873\\
$19^-$&-&5.377          &-&5.202        &-&4.111        &-&3.416        &-&3.148        &-&3.096        &-&3.057\\
$20^-$&-&5.683          &-&5.526        &-&4.329        &-&3.604        &-&3.334        &-&3.278        &-&3.242\\ \hline
\end{tabular}
\end{flushleft}

The $E\lambda$ with $\lambda=1,2,3$ transitions have been calculated by using the Eqs. (4.3-5). The negative parity states decay to the positive parity states of the partner band, by emitting E1 rays. The branching ratio
\begin{equation}
R_J=\frac{B(E1;J^-\rightarrow(J+1)^+)}{B(E1;J^-\rightarrow(J-1)^+)}.
\end{equation}
seems to be sensitive to the angular momentum as well as the specific structure of the nucleus under consideration.
In Table 8, we give this ratio for $^{226}$Ra where there are more available data for the
ground band $0^-$. Also we give the branching ratio characterizing the beta and gamma bands, respectively.

Table 8. 
{\small The branching ratios given by Eq. (5.2) are given for ground, beta and gamma bands of $^{226}$Ra. For ground band the experimental data taken from 
Ref. \cite{Smi,Coc1,Coc2} are given (left column).} 
\begin{center}
\begin{tabular}{ccccc}\hline
  $J^{\pi}$  &\multicolumn{2}{c}{g band} &{$\beta$ band} &{$\gamma$ band}\\  
  &Exp.   &Th.   &Th. &Th.\\ \hline
$1^-$ &$1.85\pm1.20$&$\hskip0.2cm$1.84&1.86&\\
$2^-$ &&&&\\
$3^-$ &$0.87\pm 0.35$&$\hskip0.2cm$ 1.12&1.14&0.20\\
$4^-$ &&&&0.92\\ 
$5^-$&&$\hskip0.2cm$ 0.94&0.96&0.49\\
$6^-$&&&&1.01\\
$7^-$&$1.79\pm 1.59$&$\hskip0.2cm$ 0.86&0.87&1.99\\
$8^-$&&&&2.70\\
$9^-$&$1.27\pm0.68$&$\hskip0.2cm$ 0.83&0.84&3.25\\
$10^-$&&&&3.71\\
$11^-$&$1.12\pm0.79$&$\hskip0.2cm$ 0.83&0.83&4.16\\
$12^-$&&&&4.60\\
$13^-$&$1.06\pm0.68$&$\hskip0.2cm$ 0.85&0.84&5.08\\
$14^-$&&&&5.57\\
$15^-$&&$\hskip0.2cm$ 0.87&0.86&6.08\\ \hline
\end{tabular}
\end{center}

The predicted ratios for the ground  $0^-$ band agree quite well with the corresponding data. It is remarkable that the ratios characterizing the $\beta^-$ band are very close to those for the ground $0^-$ band.
As for the gamma band, one notes that for odd values of angular momentum the ratio is smaller than the one characterizing the transitions of states from ground band $0^-$, carrying the same angular momentum. However, the ratio is rapidly increasing with J and already for $J^{\pi}=7^-$ it exceeds the ratio in the ground $0^-$ band. The ratio for even angular momenta $J^-$ is larger than that for
$(J-1)^-$, the curves  showing the J dependence of ratios for odd and ratios for even angular momenta are both increasing, being almost parallel with each other. In gamma band the state $J^-$ might  perform a E1 transition also to the state $J^+$. The ratio of this transition and the transition to $(J-1)^+$ is decreasing with J. For example,  it has the values 1.08, 0.35 and 0.14 for 
$J^-=5^-,15^-$ and $29^-$, respectively.  

The results for the remaining nuclei are listed in Table 9.

Table 9.  {\small The branching ratios computed using the transition operator given
by Eq.(4.2)  for $^{158}$Gd,   $^{172}$Yb,       $^{218}$Ra,
     $^{232}$Th,     $^{238}$U,      $^{238}$Pu, are compared with the
corresponding experimental data (left column). These are from Ref.\cite{Gre1}($^{158}$Gd), \cite{Blo,Gre2}($^{172}$Yb),   
\cite{Gai1,Led}($^{238}$Pu). There are no experimental data for $^{218}$Ra,
$^{232}$Th and $^{238}$U.
} 
\begin{center}
\begin{tabular}{ccccccccccccc}\hline
$J^{\pi}$&\multicolumn{2}{c}{$^{158}Gd$}   &\multicolumn{2}{c}{$^{172}Yb$} &\multicolumn{2}{c}{$^{218}Ra$}  &\multicolumn{2}{c}{$^{232}Th$}&\multicolumn{2}{c}{$^{238}U$}&\multicolumn{2}{c}{$^{238}Pu$}\\\hline
$1^-$&$1.8\pm 0.2$&$\hskip0.2cm$ 1.64&$\hskip0.2cm$$ 2.4\pm 0.6$&$\hskip0.2cm$ 1.76&$\hskip0.2cm$ -&0.88&$\hskip0.2cm$ -&1.69&$\hskip0.2cm$ -&1.86&$\hskip0.2cm$ $1.8\pm 0.05$&$\hskip0.2cm$ 1.78\\
$3^-$&$1.3\pm 0.1$&$\hskip0.2cm$ 0.84&$\hskip0.2cm$ $ 1.2\pm 0.1$&$\hskip0.2cm$ 0.99&$\hskip0.2cm$ -&0.24&$\hskip0.2cm$ -&0.91&$\hskip0.2cm$ -&1.13&$\hskip0.2cm$ $1.0\pm 0.10$&$\hskip0.2cm$ 1.02\\
$5^-$&-&$\hskip0.2cm$ 0.60&$\hskip0.2cm$ -&$\hskip0.2cm$ 0.75&$\hskip0.2cm$ -&0.18&$\hskip0.2cm$ -&0.94&$\hskip0.2cm$ -&0.66&$\hskip0.2cm$ -&$\hskip0.2cm$ 0.79\\
$7^-$&-&$\hskip0.2cm$ 0.46&$\hskip0.2cm$ -&$\hskip0.2cm$ 0.62&$\hskip0.2cm$ -&0.21&$\hskip0.2cm$ -&0.52&$\hskip0.2cm$ -&0.82&$\hskip0.2cm$ -&$\hskip0.2cm$ 0.66\\\hline
\end{tabular}
\end{center}
As shown in Ref.\cite{ButNaz}, the ratio of intrinsic dipole and quadrupole moments in $K=0$ bands can be related to the ratio of the E1 and E2 transitions:
\begin{equation}
|\frac{D_0}{Q_0}|=\left[\frac{5(I-1)}{8(2I-1)}\frac{B(E1;I\rightarrow (I-1))}{B(E2;I\rightarrow (I-2))}\right]^{1/2}.
\end{equation}
This equation is derived by using the symmetric rotor expression for the transitions involved. 
Inserting the predicted values for the B(E1) and B(E2) values, one obtains
predictions for the ratio of intrinsic moments. Possible deviations from the corresponding experimental values may reflect the deviation of the present formalism  from the symmetric rotor picture \cite{Denis}. Since the square root from the above equation is proportional to the ratio of the dipole and quadrupole effective charges $q_1/q_2$, which are not known in the present work, we fixed this ratio so that the prediction for $7^-$ fits the experimental value for $|D_0/Q_0|$.
The results for $^{226}$Ra are compared with experimental data in Table 10,
for the $K^{\pi}=0^-$ ground and beta bands.

Table 10. 
{\small The ratio of dipole and quadrupole intrinsic moments $|D_0/Q_0|$ for the ground and $\beta$ bands is
given for negative parity states of $^{226}$Ra. Theoretical predictions,
right column, are compared with the corresponding data extracted from
Ref.\cite{Wol}. Data are presented in units of $10^{-4}$ $fm^{-1}$.}
\begin{center}
\begin{tabular}{cccc}\hline
  $I^{\pi}$ &\multicolumn{2}{c}{g band} &{$\beta$ band}\\
&Exp.&Th.&Th.\\ \hline
$3^-$&& $\hskip0.2cm$ 2.41&1.52\\
$5^-$&&$\hskip0.2cm$ 2.47&1.56\\
$7^-$&&$\hskip0.2cm$ 2.52&1.59\\
$9^-$&$2.57\pm0.20$ &$\hskip0.2cm$ 2.57&1.61\\
$11^-$&$2.50\pm0.14$&$\hskip0.2cm$ 2.60&1.64\\
$13^-$&$1.96\pm0.15$&$\hskip0.2cm$ 2.64&1.66\\
$15^-$&$2.50\pm0.50$&$\hskip0.2cm$ 2.67&1.68\\
$17^-$&2.49&$\hskip0.2cm$ 2.70&1.70\\
$19^-$&&$\hskip0.2cm$ 2.73&1.73\\
$21^-$&&$\hskip0.2cm$ 2.74&1.75\\
$23^-$&&$\hskip0.2cm$ 2.67&1.78\\
$25^-$&&$\hskip0.2cm$ 2.30&1.81\\
$27^-$&&$\hskip0.2cm$ 1.38&1.85\\\hline
\end{tabular}
\end{center}

We remark a good agreement between predicted and experimental results for ground band. Although in the rotational limit the 
ratio $|D_0/Q_0|$ for ground and beta bands should be the same, in the present model the ratio for beta band is smaller 
than that for ground band. We remember that in the two bands the branching ratios  are very similar. 
The differences shown in Table 10 are determined, as we shall see a little bit later, by the fact that the B(E2) values for 
the transition $I\rightarrow(I-2)$ are  larger in beta band than the corresponding values in the ground $0^-$ band.
Before closing the discussion on the dipole transitions, some comments concerning the previous calculations are necessary. In the IBA model proposed by Iachello and Jackson [27], the high B(E1) values cannot be obtained unless a complex structure for the transition operator is considered, which allows to couple the ground state with the isovector dipole giant resonance\cite{Bren}. Although the low lying isoscalar state cannot be excited by an usual transition operator, allowing a small admixture of the giant dipole resonance in the low lying state  and assuming a "two-body" corrective term  in the transition operator, the E1 transition to the low  state $1^-$ is possible. Similar conclusion has been met in microscopic calculations of Hamamoto \cite{Hama} and Soloviov \cite{Solo} who pointed out that the dipole transition operator should be corrected with a term of octupole type which in fact is able to achieve transitions of
$\Delta N=3$ type. It is clear that in the present paper the transition operator describes transitions which, within a microscopic framework, would promote a particle to a state characterized by $\Delta N=3$. In this respect one may state that our phenomenological model is at par with the new version of IBA
\cite{Bren} as well as with the microscropic calculations \cite{Hama,Solo}.
Again, one should stress that in the present paper the dipole transition operator has a very simple structure which contrasts the transition operator used in the IBA \cite{Bren} model. This implies that the model states used in the present paper have complex structures which are suitable for accounting for the details of the e.m. properties.  

The transition operator for the E2 transitions is given by Eq.(4.1). Fixing the effective charge so that the experimental B(E2) value for the
transition $0^+\rightarrow 2^+$ is reproduced, the reduced transition probabilities in the six bands are readily obtained with Eq.(4.4). Results for $^{226}$Ra are collected 
in Tables 11-13. For the ground bands, experimental data are available. They are extracted from Ref.\cite{Wol} where data for the reduced matrix elements are given.
We note that transitions in beta bands are in general stronger than in ground bands.

\newpage
Table 11. 
{\small Reduced transition probabilities, $B(E2;J^+\rightarrow J'^+)$, for positive ground and $\beta$ bands in $^{226}$Ra, given in units of $e^2b^2$. Data are extracted from Ref.\cite{Wol}}.
\begin{center}
\begin{tabular}{p{1cm}p{2cm}p{2cm}p{2cm}p{2cm}p{2cm}p{2cm}}\hline
  $J^{\pi}$ &\multicolumn{3}{c}{$J'=J$}&\multicolumn{3}{c}{$J'=J+2$}\\
&\multicolumn{2}{c}{g band}&$\beta$ band&\multicolumn{2}{c}{g band}&$\beta$ band\\\hline
     &Exp.  &Th.   &Th.&Exp.&Th.&Th.\\ \hline
$0^+$&&&&5.14&5.14&5.96\\
$2^+$&$0.43^{+0.19}_{-0.17}$&1.46&1.70&$2.73^{+0.13}_{-0.07}$&2.68&3.10\\
$4^+$&$1.26^{+0.35}_{-0.24}$&1.33&1.54&$2.59^{+0.06}_{-0.07}$&2.43&2.79\\
$6^+$&$2.56^{+0.33}_{-0.66}$&1.30&1.50&$2.25^{+0.05}_{-0.08}$&2.37&2.70\\
$8^+$&$1.88^{+0.90}_{-0.18}$&1.29&1.48&$2.05^{+0.16}_{-0.07}$&2.38&2.69\\
$10^+$&$1.13^{+0.67}_{-0.23}$&1.29&1.47&$3.85^{+0.18}_{-0.31}$&2.42&2.72\\
$12^+$&$1.06^{+0.62}_{-0.74}$&1.28&1.46&$2.14^{+0.29}_{-0.12}$&2.48&2.77\\
$14^+$&&1.28&1.45&$2.32^{+1.07}_{-0.35}$&2.55&2.82\\
$16^+$&&1.27&1.44&$0.94^{+0.25}_{-0.20}$&2.63&2.89\\
$18^+$&&1.27&1.43&&2.74&2.97\\
$20^+$&&1.28&1.43&&2.98&3.04\\\hline
\end{tabular}
\end{center}

\newpage
Table 12. {\small The reduced transition probabilities $B(E2;J^-\to J'^-)$, for negative ground and $\beta$ bands, given in units of $e^2b^2$ for $^{226}$Ra. Data are extracted from Ref. \cite{Wol}.
The effective charge was fixed by fitting the B(E2) value for the transition $0^+\rightarrow2^+$}.
\begin{center}
\begin{tabular}{ccccccc}\hline
  $J^{\pi}$ &\multicolumn{3}{c}{$J'=J$}&\multicolumn{3}{c}{$J'=J+2$}\\
&\multicolumn{2}{c}{g band}&$\beta$ band&\multicolumn{2}{c}{g band}&$\beta$ band\\\hline
&Exp.  &Th.   &Th.   &Exp.          &Th.    &Th.\\ \hline
$1^-$&$5.74^{+0.69}_{-0.97}$&$\hskip0.2cm$ 2.05&2.37&$4.46^{+0.14}_{-0.23}$&$\hskip0.2cm$ 3.09&3.58\\
$3^-$&$2.92^{+1.42}_{-0.77}$&$\hskip0.2cm$ 1.36&1.58&$2.39^{+0.05}_{-0.12}$&$\hskip0.2cm$ 2.49&2.86\\
$5^-$&$2.11^{+0.53}_{-0.38}$&$\hskip0.2cm$ 1.32&1.52&$1.50^{+0.02}_{-0.05}$&$\hskip0.2cm$ 2.34&2.69\\
$7^-$&$2.36^{+0.56}_{-1.17}$&$\hskip0.2cm$ 1.30&1.49&$1.98^{+0.02}_{-0.12}$&$\hskip0.2cm$ 2.31&2.64\\
$9^-$&$2.41^{+1.03}_{-0.81}$&$\hskip0.2cm$ 1.29&1.48&$2.41^{+0.21}_{-0.50}$&$\hskip0.2cm$ 2.33&2.64\\
$11^-$&&$\hskip0.2cm$ 1.28&1.47&$4.35^{+0.95}_{-0.61}$&$\hskip0.2cm$ 2.38&2.68\\
$13^-$&&$\hskip0.2cm$ 1.27&1.46&$3.48^{+2.15}_{-1.58}$&$\hskip0.2cm$ 2.45&2.73\\
$15^-$&&$\hskip0.2cm$ 1.26&1.45&&$\hskip0.2cm$ 2.54&2.80\\
$17^-$&&$\hskip0.2cm$ 1.25&1.44&&$\hskip0.2cm$ 2.65&2.87\\
$19^-$&&$\hskip0.2cm$ 1.26&1.43&&$\hskip0.2cm$ 2.85&2.96\\
$21^-$&&$\hskip0.2cm$ 1.31&1.43&&$\hskip0.2cm$ 3.48&3.04\\\hline
\end{tabular}
\end{center}
\newpage
Table 13. 
{\small The reduced transition probabilities $B(E2;J^{\pm}\to J'^{\pm})$, for positive and negative $\gamma$ bands, given in units of $e^2b^2$
for $^{226}$Ra. Data are extracted from Ref.\cite{Wol}. The effective charge was determined by fitting the B(E2) value for the transition $0^+\rightarrow 2^+$.}
\begin{center}
\begin{tabular}{cccc}\hline
$J^{\pi}$ &{$J'=J$}&{$J'=J+1$}&{$J'=J+2$}\\\hline
$2^+$&0.367&0.525&0.207\\
$2^-$&0.222&0.418&0.187\\
$3^+$&0.0&0.269&0.236\\
$3^-$&0.0&0.314&0.279\\
$4^+$&0.029&0.165&0.259\\
$4^-$&0.037&0.218&0.325\\
$5^+$&0.061&0.116&0.290\\
$5^-$&0.066&0.152&0.357\\
$6^+$&0.091&0.098&0.357\\
$6^-$&0.075&0.118&0.406\\
$7^+$&0.128&0.093&0.481\\
$7^-$&0.087&0.104&0.499\\
$8^+$&0.188&0.102&0.702\\
$8^-$&0.108&0.102&0.664\\
$9^+$&0.287&0.121&1.106\\
$9^-$&0.146&0.113&0.951\\
$10^+$&0.472&0.163&1.864\\
$10^-$&0.210&0.135&1.471\\
$11^+$&0.802&0.230&3.358\\
$11^-$&0.330&0.183&2.422\\
$12^+$&1.480&0.366&6.409\\
$12^-$&0.534&0.255&4.283\\\hline

\end{tabular}
\end{center}
\newpage
The results for octupole transitions between parity partner bands are given in Tables 14, 15, for $^{226}$Ra. The effective charge $q_3$ entering the expression of the E3 transition operator is chosen so that the experimental B(E3) value for the transition $0^+_g\rightarrow3^-_g$ is reproduced. The data for the ground band are extracted from the reduced matrix elements given in Ref. \cite{Wol}.
Again the agreement with the data is fairly good.

Table 14. 
 {\small Reduced transition probabilities $B(E3;J^+\to J'^-)$ for ground and $\beta$ bands of $^{226}$Ra, given
 in units of $e^2b^3$. The effective charge was fixed so that the experimental B(E3) value for the transition
 $0^+\rightarrow 3^-$ is reproduced. Experimental data are from Ref. \cite{Wol}.}
\begin{center}
\begin{tabular}{p{.5cm}p{1.3cm}ccp{1.3cm}ccp{1.3cm}ccp{1.3cm}cc}\hline
  $J^{\pi}$ &\multicolumn{3}{c}{$J'=J-3$}&\multicolumn{3}{c}{$J'=J-1$}
&\multicolumn{3}{c}{$J'=J+1$}&\multicolumn{3}{c}{$J'=J+3$}\\\hline
&Exp.&g band&$\beta$ band&Exp.&g band&$\beta$ band&Exp.&g band&$\beta$
band&Exp.&g band&$\beta$ band\\\hline
$0^+$&&&&&&&&&&$1.17^{+0.06}_{-0.06}$&1.17&1.15\\
$2^+$&&&&$0.28^{+0.01}_{-0.03}$&0.27&0.27&$0.26^{+0.07}_{-0.07}$&0.29&0.29&$0.80^{+0.05}_{-0.05}$&0.60&0.58\\
$4^+$&$0.24^{+0.05}_{-0.02}$&0.18&0.18&$0.23^{+0.10}_{-0.04}$&0.18&0.19&$>0.32$&0.24&0.24&$0.67^{+0.04}_{-0.07}$&0.54&0.53\\
$6^+$&$0.52^{+0.20}_{-0.20}$&0.22&0.22&$0.30^{+0.09}_{-0.09}$&0.17&0.17&$0.44^{+0.52}_{-0.15}$&0.23&0.23&$0.65^{+0.04}_{-0.09}$&0.53&0.52\\
$8^+$&$>0.34$&0.24&0.24&&0.17&0.17&&0.22&0.22&$0.28^{+0.15}_{-0.44}$&0.54&0.53\\
$10^+$&&0.26&0.25&&0.17&0.17&&0.21&0.21&&0.55&0.54\\
$12^+$&&0.28&0.28&&0.17&0.17&&0.21&0.21&&0.57&0.56\\
$14^+$&&0.31&0.30&&0.17&0.17&&0.21&0.20&&0.59&0.57\\
$16^+$&&0.35&0.34&&0.17&0.17&&0.20&0.20&&0.61&0.60\\
$18^+$&&0.39&0.37&&0.18&0.17&&0.20&0.20&&0.64&0.62\\
$20^+$&&0.44&0.42&&0.18&0.18&&0.20&0.20&&0.67&0.65\\\hline
\end{tabular}
\end{center}

\newpage
Table 15. {\small Reduced transition probabilities $B(E3;J^+\to J'^-)$ for the $\gamma$ band of $^{226}$Ra, given in units of $e^2b^3$. The effective charge was fixed by fitting the B(E3) value for the transition $0^+\rightarrow 3^-$.}
\begin{center}
\begin{tabular}{p{1cm}p{2cm}p{2cm}p{2cm}p{2cm}p{2cm}p{2cm}p{2cm}}\hline
  $J^{\pi}$ &{$J-3$}&{$J-2$}&{$J-1$}&{$J$}&{$J+1$}&{$J+2$}&{$J+3$}\\\hline
$2^+$&&&&0.076&0.368&0.496&0.205\\
$3^+$&&&0.247&0.177&0.021&0.391&0.308\\
$4^+$&&0.237&0.015&0.231&0.005&0.288&0.366\\
$5^+$&0.073&0.207&0.004&0.202&0.038&0.216&0.404\\
$6^+$&0.124&0.161&0.029&0.164&0.071&0.167&0.430\\
$7^+$&0.158&0.125&0.055&0.132&0.098&0.132&0.451\\
$8^+$&0.182&0.099&0.076&0.107&0.118&0.107&0.468\\
$9^+$&0.201&0.080&0.092&0.087&0.133&0.088&0.483\\
$10^+$&0.217&0.066&0.105&0.0763&0.144&0.074&0.496\\
$11^+$&0.232&0.055&0.114&0.061&0.153&0.063&0.509\\
$12^+$&0.247&0.047&0.122&0.052&0.159&0.054&0.521\\
$13^+$&0.263&0.041&0.129&0.045&0.164&0.047&0.533\\
$14^+$&0.279&0.036&0.135&0.039&0.169&0.041&0.545\\
$15^+$&0.296&0.032&0.140&0.034&0.172&0.037&0.558\\\hline
\end{tabular}
\end{center}

\section{Conclusions}
\label{sec:level6}
In this work we developed a formalism for describing six rotational bands, three of positive and three of negative parities. These bands are organized in three pairs  $g^{\pm},\beta^{\pm},\gamma^{\pm}$, each pair being generated by projection from a quadrupole and octupole deformed state. The intrinsic functions have not good reflection symmetry and therefore the parity 
projection distinguishes between the  partner bands.

Many features seen in ground and the lowest $K^{\pi}=0^-$ bands,  customarily considered as signatures for an octupole deformation, are also seen in the other two pairs of bands called $\beta^{\pm}$ and $\gamma^{\pm}$ bands, respectively. Moreover several new properties  specific to the excited bands were pointed out.

 Partner bands intersect each other or have common J(J+1) pattern for a finite interval of angular momentum. For the angular momenta where the two energy functions  are represented by similar curves, the nucleus under consideration exhibits an octupole deformation.

The band $\beta^-$ has the same spin sequence as the band $g^-$. Also the branching ratios for the two bands are very close to each other as shown in Table 8.
However we found several specific properties for the $\beta^-$ band. Indeed, there are nuclei like $^{238}$U, $^{238}$Pu, $^{172}$Yb, where the band $\beta^-$ is lower in energy than the parity partner band, $\beta^+$. Thus, in contradistinction to the ground bands where the parity projection favors the positive parity states, for beta band this seems not to be generally true.
The negative parity beta band can intersect not only the parity partner band but also the ground negative band. This intersection is taking place in $^{238}$U already at $J=11^-$. If one compares the energy levels in the $\beta^-$ band with the experimental energies in the  $K^{\pi}=0^-$ band, one sees that the first three experimental levels, $1^-,3^-,5^-$,  are described very well by the predictions for the beta band of negative parity. To decide upon the correct interpretation of these states, data concerning the decay properties of these states are necessary.

The case of $^{172}$Yb is particularly interesting due to the the fact the band $K^{\pi}=0^-$ is highly excited. In order to describe this band as a ground band one needs a very large strength for the $\hat {N}_3$ term and a large attractive quadrupole-octupole coupling term (see the values for ${\cal B}_3$ and ${\cal B}_1$).
On the other hand such a large coupling term determines a strong lowering of the $\beta^-$ band. 
It would be interesting whether in the near future some new data about the lowest negative parity bands will appear.

To our opinion for pear shaped nuclei, the most interesting features are provided by the $\gamma^{\pm}$ bands. Indeed, the two bands are the only partner bands
having a similar spin sequence. Therefore for weakly octupole deformed nuclei, in these bands one should identify parity doublet states, i.e. states of the same
angular momenta, the same energy but of different parities. As a matter of fact
in Fig. 8 one sees that from a certain angular momentum the states of the same angular momenta from the two gamma bands are almost degenerate.

The comparison of the predicted excitation energies for the ground, $\beta^+$ and $\gamma^{\pm}$ bands with the available corresponding data is very good.
Unfortunately there are no data for the second $0^-$ band for these nuclei.
To get an idea about the quality of the agreement obtained, we mention the case of $^{232}$Th where 55 experimental energy levels are described, using 8 free parameters,  with a deviation of at most 20 KeV.

The electric transition properties of these bands were studied by using transition operators in lowest order in quadrupole and octupole bosons. The theoretical  E1 branching ratios, ratios of dipole and quadrupole intrinsic moments,
quadrupole intraband transitions as well as octupole transitions between parity partner bands have been compared with existing data for ground bands in
 $^{226}$Ra. The agreement is very good. Also the predictions for excited pairs, beta and gamma, are considered and comments upon how the results compare with the similar results in ground bands are given. The E1 branching ratios are listed for few states of the ground negative parity band for all nuclei considered in the present work.

As a final conclusion, the present paper is the first one in the literature which treats in an unified fashion the positive and negative parity bands. Indeed,
in contrast to the previous paper where one describes on an equal footing the ground and the lowest $K^{\pi}=0^-$ bands here one considers in addition two pairs of excited bands. The agreement with the existing data encourages us to
enlarge the number of  pairs of parity partner bands. 

{\bf Acknowledgements.} Part of this work was supported by CNCSIS under contract A918/2001, Romania, and part by the Humboldt Foundation.

\section{Appendix A}
\renewcommand{\theequation}{A.\arabic{equation}}
\setcounter{equation}{0}
\label{sec:level A}
The diagonal matrix elements of the model Hamiltonian in the basis of projected states can be written in a compact form:
\begin{eqnarray}
\langle\varphi_J^{(i,k)}|H|\varphi_J^{(i,k)}\rangle&=&{\sum_{J_2,J_3}}^{(i)}X_{J,k}^{J_2J_3}
\{({\cal A}_1+{\cal B}_1y_3\frac{{\cal I}_{J_3}^{(k)'}}{{\cal I}_{J_3}^{(k)}})
\langle\varphi_{J_2}^{(i)}|22\hat{N}_2+5\Omega^{\dagger}_{\beta'}\Omega_{\beta'}|\varphi_{J_2}^{(i)}\rangle
\nonumber\\
&&+\delta_{(i,\beta)}({\cal A}_2+{\cal B}_2y_3\frac{{{\cal I}^{(k)}_{J_3}}^{\prime}}
{{\cal I}^{(k)}_{J_3}})\langle\varphi^{(\beta)}_{J_2}|\Omega^{\dag}_{\beta}\Omega_{\beta}|\varphi^{(\beta)}_{J_2}\rangle
+{\cal B}_3y_3\frac{{\cal I}_{J_3}^{(k)'}}{{\cal I}_{J_3}^{(k)}}\nonumber\\
&&+
{\cal A}_{(J23)}\frac{1}{2}\left[J(J+1)-J_2(J_2+1)-J_3(J_3+1)\right]\}
+{\cal A}_JJ(J+1)\equiv H^{(k)}_{ii;J},\nonumber\\
&&i=g,\beta,\gamma,~k=\pm
\end{eqnarray}
where the coefficients X are given by:
\begin{equation}
^{(i)}X_{J,k}^{J_2J_3}=(N_J^{(i,k)})^{2}(N_{J_3}^{(k)}N_{J_2}^{(i)})^{-2}
(C_{0\hskip0.25cmK_i\hskip0.2cmK_i}^{J_3\hskip0.1cm J_2\hskip0.2cm J})^2,~~
k=\pm,~ i=g, \beta, \gamma.
\end{equation}
The matrix elements between the CSM projected states are given analytically
in Ref.\cite{Rad11}.
The off-diagonal matrix elements are between the states of gamma and ground bands:
\begin{eqnarray}
\langle \varphi^{(g,k)}_{JM}|H|\varphi^{(\gamma,k)}_{JM}\rangle &=&
{\cal N}^{(g,k)}_J{\cal N}^{(\gamma,k)}_J\sum_{J_2,J_3}\left(N^{(k)}_{J_3}\right)^{-2}\left(N^{(g)}_{J_2}
N^{(\gamma,k)}_{J_2}\right)^{-1}C^{J_2~J_3~J}_{0~~~0~~~0}C^{J_2~J_3~J}_{0~~~2~~~2}\nonumber\\
&&\times\left({\cal A}_1+{\cal B}_1y_3\frac{{{\cal I}^{(k)}_{J_3}}^{\prime}}{{\cal I}^{(k)}_{J_3}}\right)
\langle\varphi^{(g)}_{J_2}|22{\hat N}+
5\Omega^{\dag}_{\beta'}\Omega_{\beta'}|\varphi^{(\gamma)}_{J_2}\rangle
\equiv H^{(k)}_{g\gamma;J}.
\end{eqnarray}
The matrix elements involved in the expression of the B(E1) value are given in Appendix B.

\section{Appendix B}
\renewcommand{\theequation}{B.\arabic{equation}}
\setcounter{equation}{0}
\label{sec:levelB}
Here the reduced matrix elements of the dipole transition operator are expressed in terms of the norms and the matrix elements between the CSM states which were previously calculated \cite{Rad11}. The diagonal matrix elements can be written in a symmetric form:
\begin{eqnarray}
\langle\varphi_I^{(g,-)}||T_1||\varphi_J^{(g,+)}\rangle&=&f\hat{1}\hat{J}{\cal N}_I^{(g,-)}{\cal N}_J^{(g,+)}
\sum_{J_2,J_2',J_3,J'}\left(N_{J_2'}^{(g)}N_{J_2}^{(g)}\right)^{-1}\hat{J'_2}\hat{J'}
\\
&\times&\langle\varphi_{J'_2}^{(g)}||b^{\dagger}_{2}+b_2||\varphi_{J_2}^{(g)}\rangle
\left[F_{IJ}^{J_2J_2'J_3J'}\left(N_{J_3}^{(+)}\right)^{-2}-F_{JI}^{J_2'J_2J_3J'}\left(N_{J_3}^{(-)}\right)^{-2}\right],\nonumber
\end{eqnarray}
\begin{eqnarray}
\langle\varphi_I^{(\beta,-)}||T_1||\varphi_J^{(\beta,+)}\rangle&=&f\hat{1}\hat{J}{\cal N}_I^{(\beta,-)}{\cal N}_J^{(\beta,+)}
\sum_{J_2,J_2',J_3,J'}\left(N_{J_2}^{(\beta)}N_{J_2'}^{(\beta)}\right)^{-1}\hat{J'_2}\hat{J'}
\\
&\times&\langle\varphi_{J'_2}^{(\beta)}||b^{\dagger}_{2}+b_2||\varphi_{J_2}^{(\beta)}\rangle
\left[F_{IJ}^{J_2J_2'J_3J'}\left(N_{J_3}^{(+)}\right)^{-2}-F_{JI}^{J_2'J_2J_3J'}\left(N_{J_3}^{(-)}\right)^{-2}\right],\nonumber
\end{eqnarray}
\begin{eqnarray}
\langle\varphi_I^{(\gamma,-)}||T_1||\varphi_J^{(\gamma,+)}\rangle&=&f\hat{1}\hat{J}{\cal N}_I^{(\gamma,-)}{\cal N}_J^{(\gamma,+)}
\sum_{J_2,J_2',J_3,J'}\left(N_{J_2}^{(\gamma)}N_{J_2'}^{(\gamma)}\right)^{-1}\hat{J'_2}\hat{J'}C_{0\hskip0.27cm2\hskip0.25cm2}^{J_3\hskip0.1cm J_2'\hskip0.2cmJ'}
\\
&\times&\langle\varphi_{J'_2}^{(\gamma)}||b^{\dagger}_{2}+b_2||\varphi_{J_2}^{(\gamma)}\rangle
\left[F_{\gamma
    \gamma;IJ}^{J_2J_2'J_3J'}\left(N_{J_3}^{(+)}\right)^{-2}+(-)^{I-J+J_2'-J_2}F_{\gamma \gamma;JI}^{J_2J_2'J_3J'}\left(N_{J_3}^{(-)}\right)^{-2}\right].
\nonumber
\end{eqnarray}
The coefficients F have the expressions:
\begin{eqnarray}
F_{JI}^{J_2J_2'J_3J'}&=&C_{0\hskip0.27cm0\hskip0.3cm0}^{J_3\hskip0.1cm
  J_2\hskip0.2cm J} C_{0\hskip0.27cm0\hskip0.3cm0}^{J_3\hskip0.1cm
  J_2'\hskip0.2cm J'} 
C_{0\hskip0.27cm0\hskip0.25cm0}^{J'\hskip0.1cm 3\hskip0.2cm I}
  W(J'J_32J_2;J'_2J) W(J'3J1;I2),
\nonumber\\
F_{\gamma \gamma;JI}^{J_2J_2'J_3J'}&=&C_{0\hskip0.27cm2\hskip0.3cm2}^{J_3\hskip0.1cm
  J_2\hskip0.2cm J}  C_{2\hskip0.27cm0\hskip0.3cm2}^{J_3'\hskip0.1cm
  3\hskip0.2cm I} W(J'J_32J_2;J_2'J) W(J'3J1;I2).
\end{eqnarray}
Standard notation for the Racah coefficients W(a,b,c,d;ef) has been used.
The matrix elements between the ground and gamma  bands states have the following expressions:

\begin{eqnarray}
\langle\varphi_I^{(\gamma,-)}||T_1||\varphi_J^{(g,+)}\rangle&=&f\hat{1}\hat{J}{\cal N}_I^{(\gamma,-)}{\cal N}_J^{(g,+)}
\sum_{J_2,J_2',J_3,J'} \left(N_{J_2'}^{(\gamma)}N_{J_2}^{(g)}\right)^{-1}\hat{J'_2}\hat{J'}
\langle\varphi_{J'_2}^{(\gamma)}||b^{\dagger}_{2}+b_2||\varphi_{J_2}^{(g)}\rangle[(-)^{J'-I}C_{2\hskip0.27cm0\hskip0.25cm2}^{J'\hskip0.1cm 3\hskip0.2cm
    I}\nonumber \\
&\times&
   C_{0\hskip0.27cm0\hskip0.25cm0}^{J_3\hskip0.1cm J_2\hskip0.2cm
    J}C_{0\hskip0.27cm2\hskip0.25cm2}^{J_3\hskip0.1cm J'_2\hskip0.2cm
    J'}\times W(J'J_32J_2;J'_2J) \times W(J'3J1;I2)
  \left(N_{J_3}^{(+)}\right)^{-2}
\\
&+&(-)^{J'-J}C_{0\hskip0.27cm0\hskip0.25cm0}^{J'\hskip0.1cm 3\hskip0.2cm
  J}C_{0\hskip0.27cm0\hskip0.25cm0}^{J_3\hskip0.1cm J_2\hskip0.2cmJ'}
  C_{0\hskip0.27cm2\hskip0.25cm2}^{J_3\hskip0.1cmJ_2'\hskip0.2cmI} 
  W(IJ_32J_2;J_2'J') W(J3I2;J'1)\left (N_{J_3}^{(-)}\right )^{-2}],\nonumber
\end{eqnarray}
\begin{eqnarray}
\langle\varphi_I^{(g,-)}||T_1||\varphi_J^{(\gamma,+)}\rangle&=&f\hat{1}\hat{J}{\cal N}_I^{(g,-)}{\cal N}_J^{(\gamma,+)}
\sum_{J_2,J_2',J_3,J'} \left(N_{J_2'}^{(\gamma)}N_{J_2}^{(g)}\right)^{-1}\hat{J'_2}\hat{J'}
\langle\varphi_{J'_2}^{(g)}||b^{\dagger}_{2}+b_2||\varphi_{J_2}^{(\gamma)}\rangle[C_{0\hskip0.27cm0\hskip0.25cm0}^{J'\hskip0.1cm 3\hskip0.2cm
    I}\nonumber \\
&\times&
   C_{0\hskip0.27cm0\hskip0.25cm0}^{J_3\hskip0.1cm J'_2\hskip0.2cm
    J'}C_{0\hskip0.27cm2\hskip0.25cm2}^{J_3\hskip0.1cm J_2\hskip0.2cm
    J}\times W(J'J_32J_2;J'_2J) \times W(J'3J1;I2)
  \left(N_{J_3}^{(+)}\right)^{-2}
\\
&+&(-)^{J'-J}C_{2\hskip0.27cm0\hskip0.25cm2}^{J'\hskip0.1cm 3\hskip0.2cm
  J}C_{0\hskip0.27cm2\hskip0.25cm2}^{J_3\hskip0.1cm J_2\hskip0.2cmJ'}
  C_{0\hskip0.27cm0\hskip0.25cm0}^{J_3\hskip0.1cmJ_2'\hskip0.2cmI} 
  W(IJ_32J_2;J'_2J') W(J3I2;J'1)\left (N_{J_3}^{(-)}\right )^{-2}],\nonumber
\end{eqnarray}

\section{Appendix C}
\renewcommand{\theequation}{C.\arabic{equation}}
\setcounter{equation}{0}
\label{sec:level C}
The reduced matrix elements of the octupole transition operator can be expressed in terms of norms of the states involved in the given transition.
\begin{equation}
\langle\varphi_{J}^{(g,+)}||Q_{3}||\varphi_{J'}^{(g,-)}\rangle=
C_{0\hskip0.27cm0\hskip0.3cm0}^{J'\hskip0.1cm3\hskip0.25cmJ} f \left(\frac{{\cal N}_J^{(g,+)}}{{\cal N}_{J'}^{(g,-)}}+
\frac{{\cal N}_{J'}^{(g,-)}}{{\cal N}_J^{(g,+)}} \frac{2J'+1}{2J+1}\right),
\end{equation}

\begin{equation}
\langle\varphi_{J}^{(\gamma,+)}||Q_3||\varphi_{J'}^{(\gamma,-)}\rangle=f
C_{0\hskip0.27cm0\hskip0.3cm0}^{J'\hskip0.1cm3\hskip0.25cmJ} \left(\frac{{\cal N}_J^{(\gamma,+)}}{{\cal N}_{J'}^{(\gamma,-)}}+
\frac{{\cal N}_{J'}^{(\gamma,-)}}{{\cal N}_J^{(\gamma,+)}} \frac{2J'+1}{2J+1}\right),
\end{equation}

\begin{equation}
\langle\varphi_{J}^{(\beta,+)}||Q_3||\varphi_{J'}^{(\beta,-)}\rangle=f
C_{0\hskip0.27cm0\hskip0.3cm0}^{J'\hskip0.1cm3\hskip0.25cmJ} \left(\frac{{\cal N}_J^{(\beta,+)}}{{\cal N}_{J'}^{(\beta,-)}}+
\frac{{\cal N}_{J'}^{(\beta,-)}}{{\cal N}_J^{(\beta,+)}} \frac{2J'+1}{2J+1}\right).
\end{equation}
If the states participating to the octupole transitions are mixtures of ground and gamma states the matrix elements are:

\begin{eqnarray}
\langle\Psi_J^{(g,+)}||Q_3||\Psi_{J'}^{(g,-)}\rangle&=&X_J^{(g,+)}X_{J'}^{(g,-)}\langle\varphi_J^{(g,+)}||Q_3||\varphi_{J'}^{(g,-)}\rangle
\nonumber\\
&&+Y_J^{(\gamma,+)}Y_{J'}^{(\gamma,-)}\langle\varphi_J^{(\gamma,+)}||Q_3||\varphi_{J'}^{(\gamma,-)}\rangle, 
\end{eqnarray}

\begin{eqnarray}
\langle\Psi_J^{(\gamma,+)}||Q_3||\Psi_{J'}^{(\gamma,-)}\rangle&=&X_J^{(\gamma,+)}X_{J'}^{(\gamma,-)}\langle\varphi_J^{(g,+)}||Q_3||\varphi_{J'}^{(g,-)}\rangle
\nonumber\\
&&+Y_J^{(\gamma,+)}Y_{J'}^{(\gamma,-)}\langle\varphi_J^{(\gamma,+)}||Q_3||\varphi_{J'}^{(\gamma,-)}\rangle
\end{eqnarray}

\begin{figure}[h]
\centerline{\psfig{figure=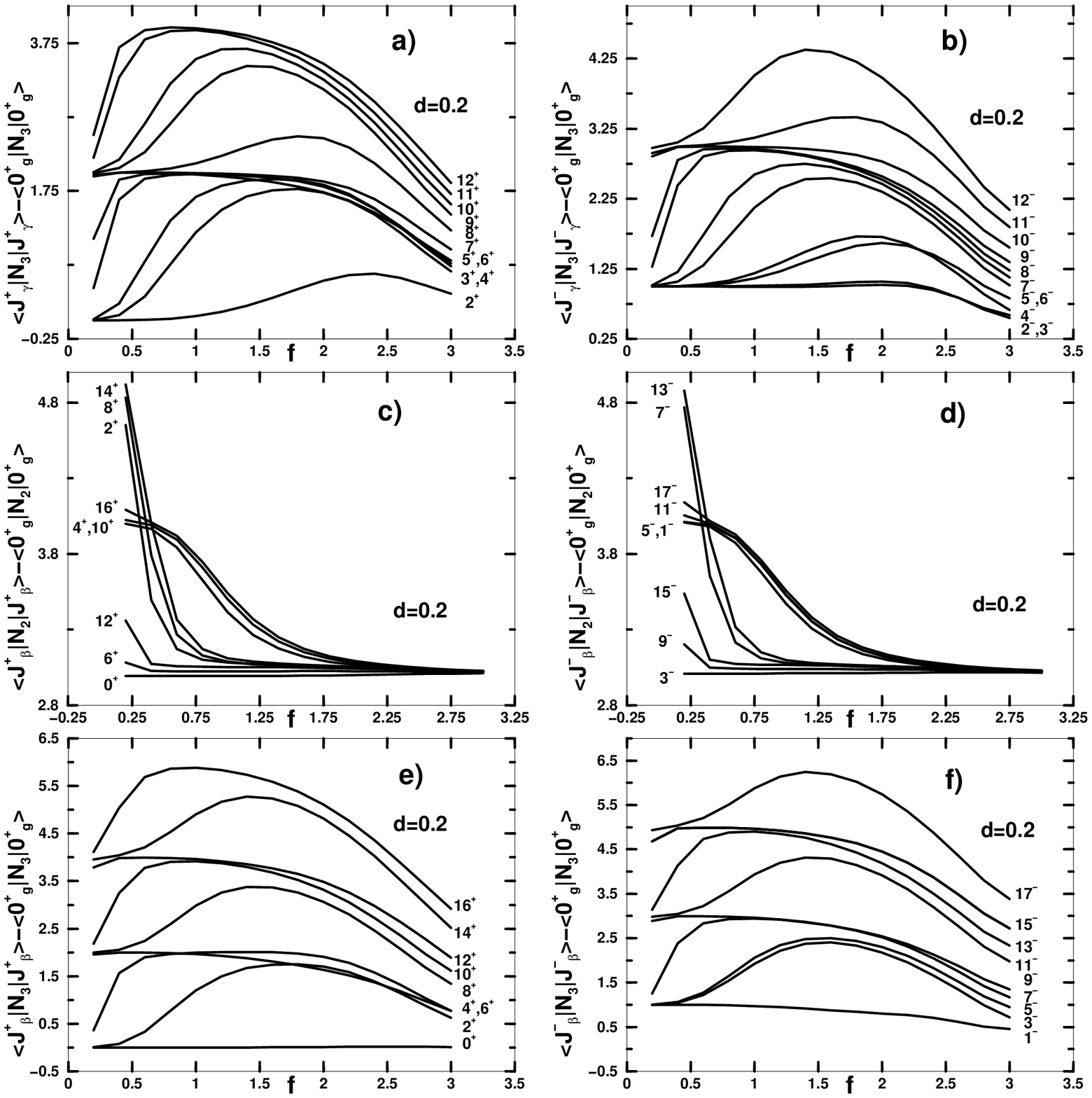,width=14cm,bbllx=2cm,%
bblly=5cm,bburx=20cm,bbury=26cm,angle=0} }
\vskip4cm
\caption{Normalised average values, for $N_2$ in beta band states (c) and d))
and for $N_3$ in gamma (a) and b)) and beta (e) and f)) band states,
are given as function of f for d=0.2. }
\label{Fig. 1}
\end{figure}
\clearpage

\begin{figure}[h]
\centerline{\psfig{figure=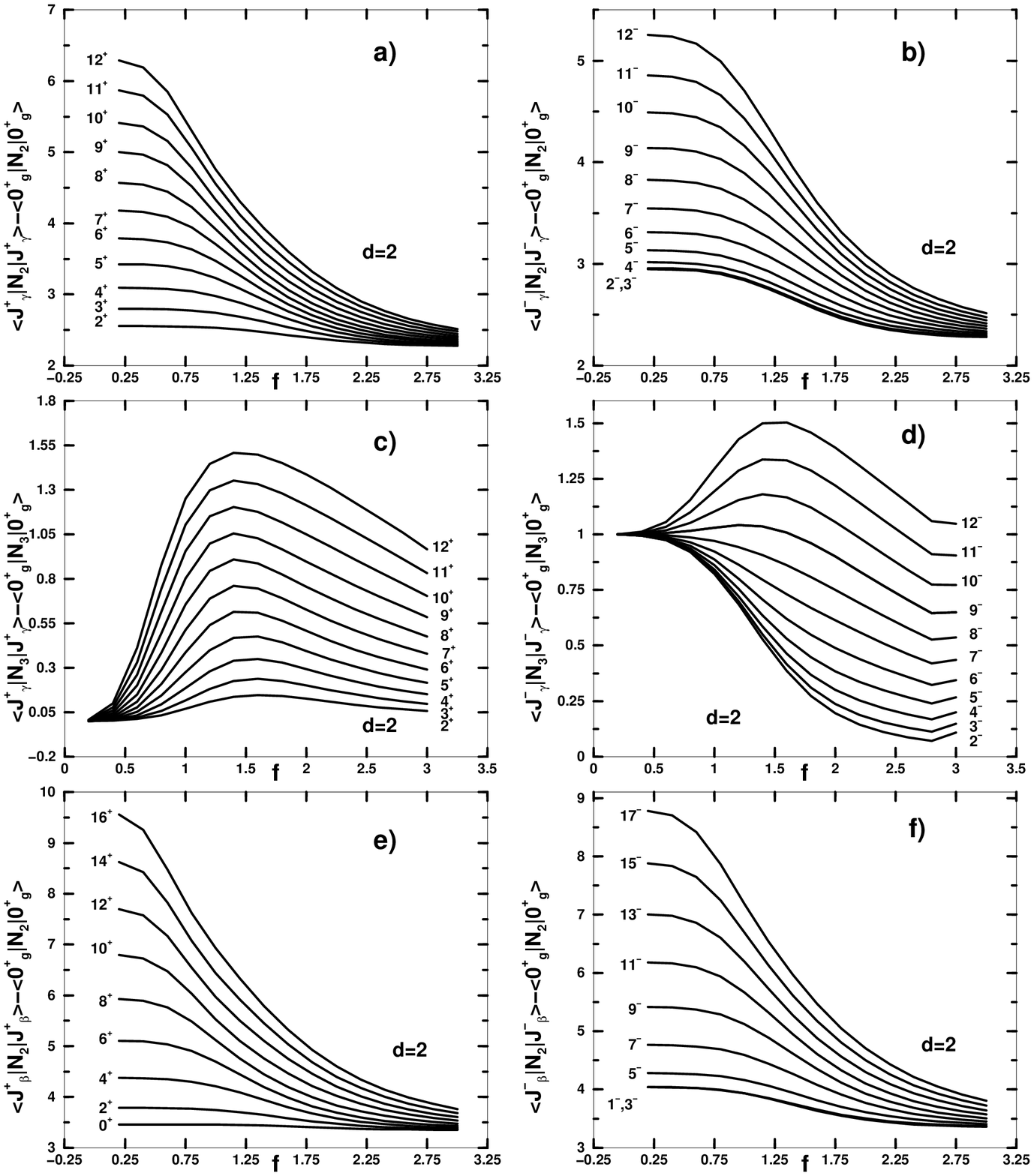,width=14cm,bbllx=2cm,%
bblly=5cm,bburx=20cm,bbury=26cm,angle=0} }
\vskip4cm
\caption{Normalised average values for $N_2$, in beta (e) and f)) and gamma
band states (a) and b))
and for $N_3$ in gamma (c) and d)) and  band states,
are given as function of f for d=2. }
\label{Fig. 2}
\end{figure}
\clearpage
\begin{figure}[h]
\centerline{\psfig{figure=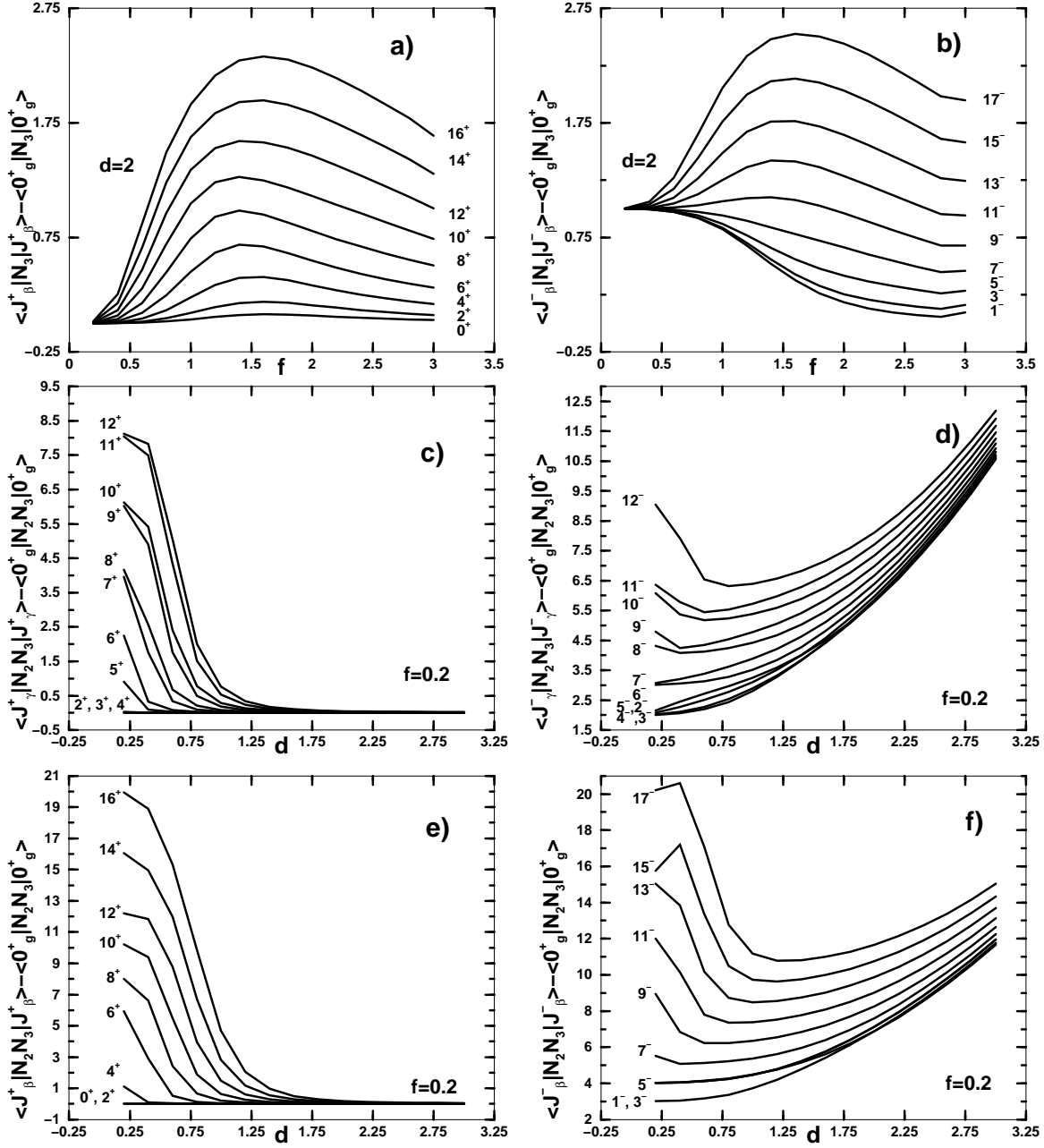,width=14cm,bbllx=2cm,%
bblly=5cm,bburx=20cm,bbury=26cm,angle=0} }
\vskip4cm
\caption{Normalised average values for $N_3$ in beta (a) and b))
band states are given as function of f for d=2. Also the averages 
 for $N_2N_3$ in gamma (c) and d)) and beta (e) and f)) band states
are given as function of d for f=0.2. }
\label{Fig. 3}
\end{figure}
\clearpage

\begin{figure}[h]
\centerline{\psfig{figure=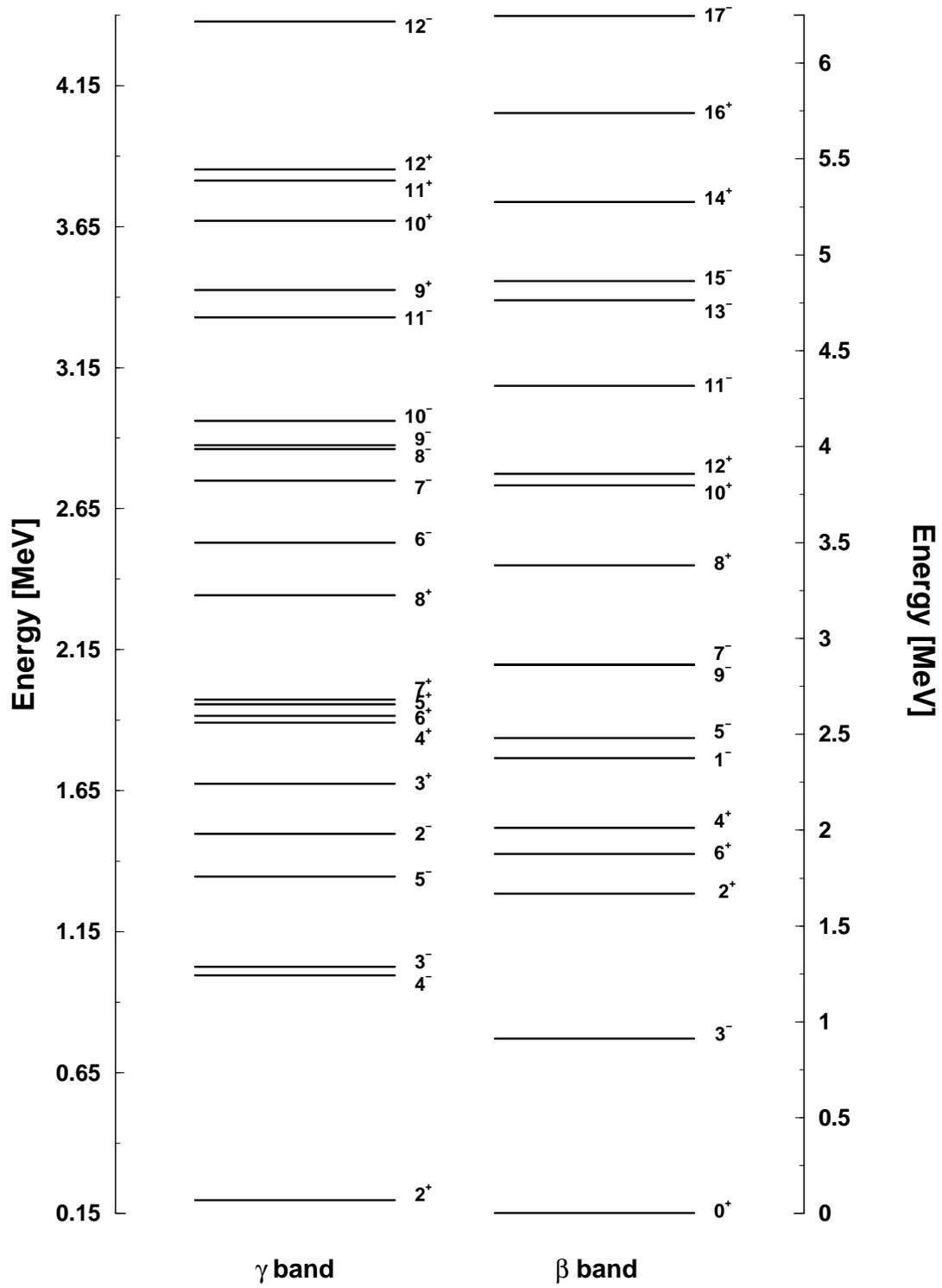,width=14cm,bbllx=2cm,%
bblly=5cm,bburx=20cm,bbury=26cm,angle=0} }
\vskip4cm
\caption{The shell structure in gamma (left) and beta bands shown in the spectrum of the $N_3$ operator for d=0.2 and f=1.4.}
\label{Fig. 4}
\end{figure}
\clearpage
\begin{figure}[h]
\centerline{\psfig{figure=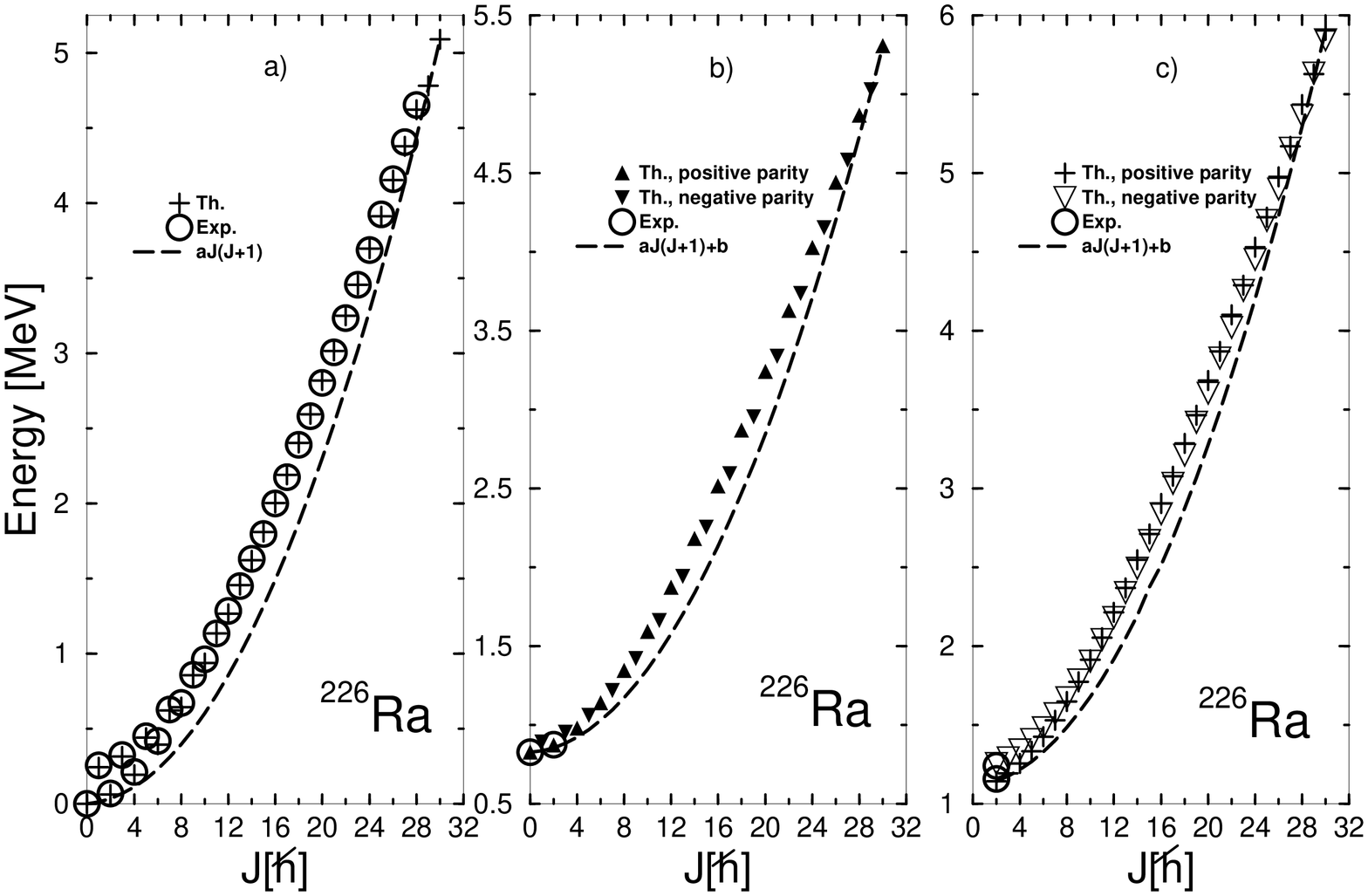,width=12cm,bbllx=1cm,%
bblly=5cm,bburx=20cm,bbury=26cm,angle=-90} }
\vskip4cm
\caption{Predicted energies for six rotational bands, $g^{(\pm)},\beta^{(\pm)},
\gamma^{(\pm)}$ are compared with the corresponding experimental data for 
$^{226}$Ra. With dashed line is represented the parabola $aJ(J+1)+b$ reaching the first and last energy in the positive parity bands.}
\label{Fig. 5}
\end{figure}
\clearpage

\begin{figure}[h]
\centerline{\psfig{figure=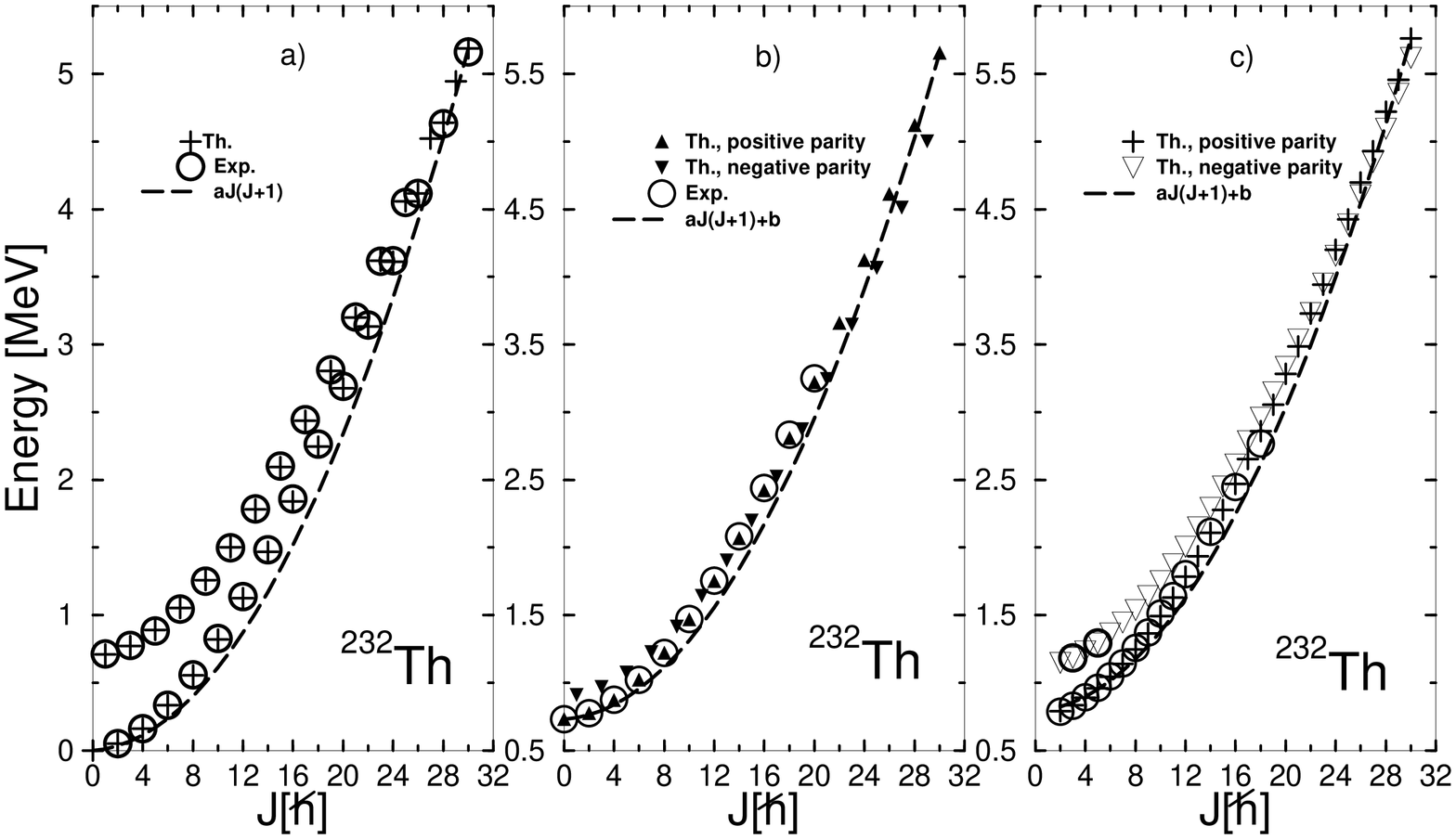,width=12cm,bbllx=1cm,%
bblly=5cm,bburx=20cm,bbury=26cm,angle=-90} }
\vskip4cm
\caption{Predicted energies for six rotational bands, $g^{(\pm)},\beta^{(\pm)},
\gamma^{(\pm)}$ are compared with the corresponding experimental data for 
$^{232}$Th. With dashed line is represented the parabola $aJ(J+1)+b$ reaching the first and last energy in the positive parity bands.}
\label{Fig. 6}
\end{figure}
\clearpage

\begin{figure}[h]
\centerline{\psfig{figure=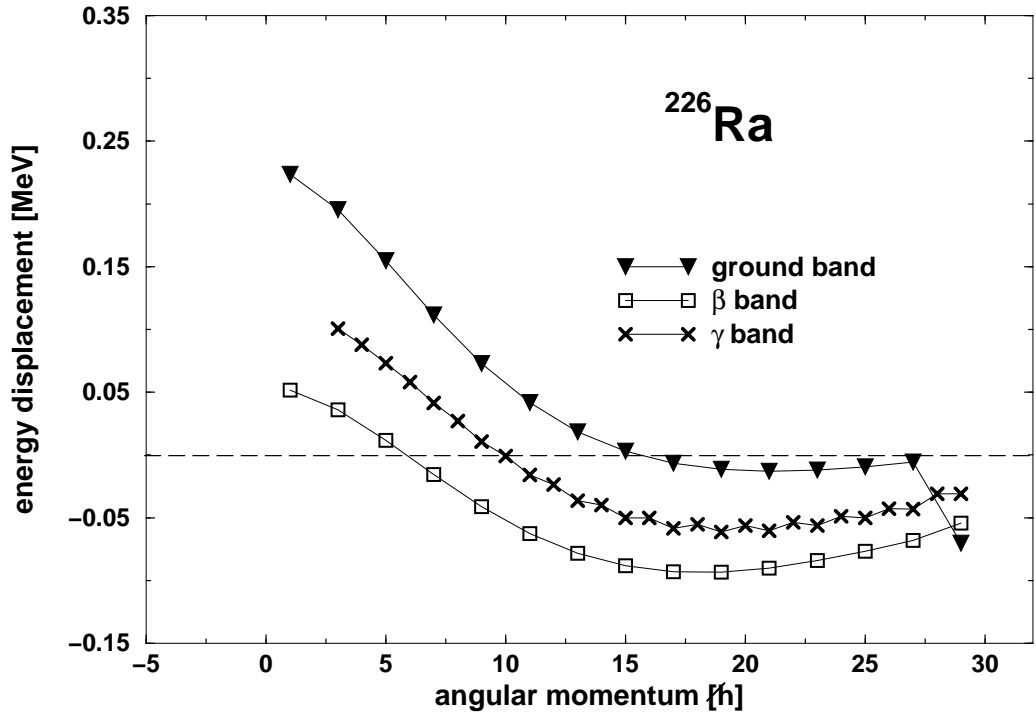,width=12cm,bbllx=1cm,%
bblly=5cm,bburx=20cm,bbury=26cm,angle=-90} }
\vskip4cm
\caption{Energy displacement functions, for the three pairs of bands, ground, beta and gamma, are plotted as function of angular momentum for $^{226}$Ra. }
\label{Fig. 7}
\end{figure}
\clearpage

\begin{figure}[h]
\centerline{\psfig{figure=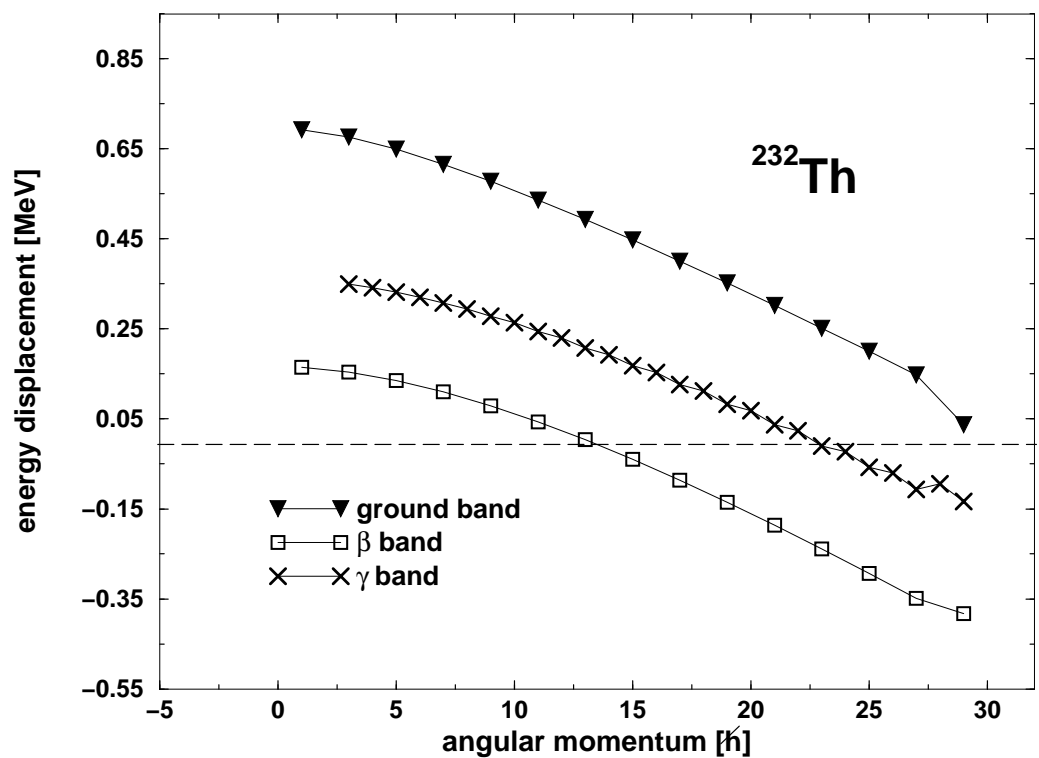,width=12cm,bbllx=1cm,%
bblly=5cm,bburx=20cm,bbury=26cm,angle=-90} }
\vskip4cm
\caption{The same as in Fig. 7 but for $^{232}$Th.}
\label{Fig. 8}
\end{figure}
\clearpage

\end{document}